\documentclass[fleqn,usenatbib,useAMS]{mnras}

\usepackage{graphicx}	
\usepackage{amsmath}	
\usepackage{amssymb}	
\usepackage{multicol}        
\usepackage{bm}		
\usepackage{pdflscape}	
\usepackage[T1]{fontenc}
\usepackage{ae,aecompl}
\usepackage{newtxtext,newtxmath}

\title[Neutron Star Merger Nebula]{Nebular Emission from Lanthanide-rich Ejecta of Neutron Star Merger}

\author[K. Hotokezaka et al.]{Kenta Hotokezaka,$^{1,2}$\thanks{E-mail: kentah@g.ecc.u-tokyo.ac.jp}
Masaomi Tanaka,$^3$
Daiji Kato,$^{4,5}$
Gediminas Gaigalas$^6$\\
$^1$Research Center for the Early Universe, Graduate School of Science, University of Tokyo, Bunkyo, Tokyo 113-0033, Japan\\
$^2$Kavli IPMU (WPI), UTIAS, The University of Tokyo, Kashiwa, Chiba 277-8583, Japan\\u University, Aoba, Sendai 980-8578, Japan\\
$^3$Astronomical Institute, Tohoku University, Aoba, Sendai 980-8578, Japan\\
$^4$National Institute for Fusion Science, 322-6 Oroshi-cho, Toki 509-5292, Japan\\
$^5$Department of Advanced Energy Engineering Science, Kyushu University, Kasuga, Fukuoka 816-8580, Japan\\
$^6$Institute of Theoretical Physics and Astronomy, Vilnius University, Saul\.{e}tekio Ave. 3, Vilnius, Lithuania
}
\pubyear{2020}

\begin{document}
\label{firstpage}
\pagerange{\pageref{firstpage}--\pageref{lastpage}}
\maketitle

\begin{abstract}
The nebular phase of lanthanide-rich ejecta of a neutron star merger (NSM)  is studied by using  a one-zone model, in which the atomic properties are represented by a single species, neodymium (Nd). Under the assumption that $\beta$-decay of $r$-process nuclei is  the heat and ionization source,
we solve the ionization and thermal balance of the ejecta under non-local thermodynamic equilibrium. 
The atomic data including energy levels, radiative transition rates, collision strengths, and recombination rate coefficients, are obtained by using atomic structure codes, \texttt{GRASP2K} and \texttt{HULLAC}.
We find that both permitted and forbidden lines roughly equally contribute to the cooling rate of Nd II and Nd III at the nebular temperatures.
We show that the kinetic temperature and ionization degree increase with time in the early stage of  the nebular phase while these quantities become approximately independent of time after the thermalization break of the heating rate because the processes relevant to the ionization and thermalization balance are attributed to two-body collision between electrons and ions at later times. 
As a result, in spite of the rapid decline of the luminosity, the shape of the emergent spectrum does not change significantly with time after the break. We show that the emission-line nebular spectrum of the pure Nd ejecta consists of a broad structure from  $0.5\,{\rm \mu m}$ to  $20\,{\rm \mu m}$   with  two  distinct  peaks  around $1\,{\rm \mu m}$   and $10\,{\rm \mu m}$.
\end{abstract}

\section{Introduction} \label{sec:intro}
Neutron star mergers (NSMs) have been considered as the sites of $r$-process nucleosynthesis \citep{lattimer1974ApJ}. In August 2017, the LIGO/Virgo Collaboration (LVC) discovered the first NSM, GW170817, which was accompanied by 
radiation across the entire electromagnetic spectrum \citep{Abbott2017ApJ,Nakar2020PhR,Margutti2020}. In particular, the spectrum and light curve of the uv-optical-infrared counterpart referred to as `kilonova' or `macronova' indicate that a copious amount of $r$-process elements is produced in this event \citep{Andreoni2017PASA,Arcavi2017Natur,Coulter2017,Cowperthwaite2017,Drout2017,Evans2017,Kasliwal2017,Pian2017,smartt2017Natur,tanvir2017ApJ,Utsumi2017PASJ}. The amount of the produced $r$-process elements and the event rate estimated from  GW170817 suggest that NSMs could provide all the $r$-process elements in the Galaxy (e.g. \citealt{Hotokezaka2018IJMPD,Rosswog2018A&A}).

Lanthanide ions have unique optical properties, which enhance the opacity of the NSM ejecta material \citep{barnes2013ApJ,kasen2013ApJ,tanaka2013ApJ,Wollaeger,Bulla2019MNRAS,Barnes2020}. Thus, the existence of lanthanide ions imprints observable signatures in kilonova light curves and spectra.
In fact, the late-time spectra of GW170817 peaking around the near infrared (nIR) band implies that  lanthanides exist in the GW170817 ejecta \citep{Kasen2017,Tanaka2017PASJ}. At the same time,
the light curve rises on a short time  scale of $\sim 0.5$ day, suggesting that there is a lanthanide-free ejecta component.   Various models have been proposed to explain  the coexistence of lanthanide-rich and free components in the GW170817 ejecta \citep{Kasen2017,Tanaka2017,Villar2017,Waxman2018MNRAS,Shibata2017,Perego2017,kawaguchi2018,hotokezaka2020ApJ}.

Recently, \cite{Watson2019Natur} analyzed the observed spectra of GW170817 with an assumption that the spectra from $\sim 1$ to $5$ day  consist of a single temperature blackbody with structure produced by atomic transitions. 
They found that the main structure of the spectra is consistent with the P Cygni profiles produced by Sr II doublet $4078,\,4215{\AA}$ and triplet $10327,10037,10915{\AA}$.
Interestingly, the Sr II lines are one of a few prominent features in the synthetic spectra  in \cite{tanaka2013ApJ}.  \cite{Perego2020} found an alternative interpretation that this spectral structure can be attributed to the He lines while they concluded this interpretation is less likely. \cite{Gillanders2021} seek the signatures of platinum and gold. However, they did not find  such signatures in the early spectra of the GW170817 kilonova.

The question is now - can  more elements be identified from kilonova observations? The direct detection of nuclear $\gamma$-rays can be one of the most robust identifications of radioactive isotopes \citep{hotokezaka2016MNRAS,Li2019ApJ,wu2019ApJb,korobkin2020ApJ}. 
However, such measurements are very challenging and only weak upper limits were put by {\it NuSTAR} in GW170817 \citep{Evans2017}.

Here we consider the nebular phase of kilonovae, where the emergent spectrum is dominated by emission lines, and hence, spectroscopic observations may enable  to identify the elements produced in NSMs. Since the slower ejecta component can be observed in the nebular phase than the earlier phases one can expect that the Doppler broadening of lines is weaker so that the spectral structure arising from individual lines may be more pronounced.  
In GW170817, the {\it Spitzer} Space Telescope detected the late-time nebular emission at $4.5\,{\rm \mu m}$
and put upper limits at $3.6\,{\rm \mu m}$  \citep{Kasliwal2019MNRAS,Villar2018ApJ}, suggesting that a fraction of the luminosity of the nebula  is radiated in infrared with a peculiar spectral shape. 



The primary goal of this paper is to address  the evolution of thermodynamic quantities and the emerging spectral shape of lanthanide-rich NSM nebulae. {The early works on the nebular modelings of kilonovae assume  local thermodynamic equilibrium (LTE) for ionization and level population \citep{Waxman2018MNRAS,Gillanders2021}. However, the non-LTE effects are crucial for the late-time nebular modelings.}  Here we develop a  NSM nebula model under non-LTE by following the studies of supernova (SN) nebular emission \citep{Axelrod1980,Fransson1989ApJ,Ruiz-Lapuente1992ApJ,Mazzali2006ApJ,Maeda2006ApJ,Boty2018ApJ}.
The paper is organized as follows. In \S \ref{sec:heat}, we describe the heating and ionization rates due to $\beta$-decay of $r$-process nuclei. In \S \ref{sec:eqs}, we describe the  equations and several approximations used in the modeling. In \S \ref{sec:atom},
we show the atomic data obtained by using the atomic codes. In \S \ref{sec:evolve}, we apply our model to a lanthanide-rich NSM nebula and show the time evolution of temperature, ion abundances, and emission spectra.  We conclude and discuss our study in \S \ref{sec:conc}.

\section{Time scale, radioactive heat, and ionization} \label{sec:heat}
Calculating the nebular emission generally requires radiative transfer computations under non-LTE.
However, such computations for NSM nebulae demand a lot of effort.
As a first step, we focus here on the nebular phase where the following conditions are satisfied:  (i) the ejecta is optically thin  and (ii) the recombination and cooling times are shorter than the dynamical time. The former allows us to simplify the treatment of radiative transfer and the latter allows to use the steady-state approximation. 

   The optical depth of the NSM ejecta is estimated by
\begin{eqnarray}
\tau \approx \frac{\kappa M_{\rm ej}}{4\pi v_0^2t^2},
\end{eqnarray}
 where $M_{\rm ej}$ is the ejecta mass and  $\kappa$ is the ejecta opacity. Here,
we assume a homologous expansion of the ejecta, i.e.,   
 $\rho(t,v) \propto t^{-3}v^{-\alpha}$ between $v_0$ and $v_1$, 
 where $\rho$ is the ejecta density, $t$ is time since merger, $v_0$ and $v_1$ are the minimum and maximum expansion velocities, and $\alpha$ describes the velocity profile of the ejecta. In this paper, we consider spherical symmetric ejecta for simplicity.
 
\begin{figure}
\centering
\includegraphics[scale=0.5]{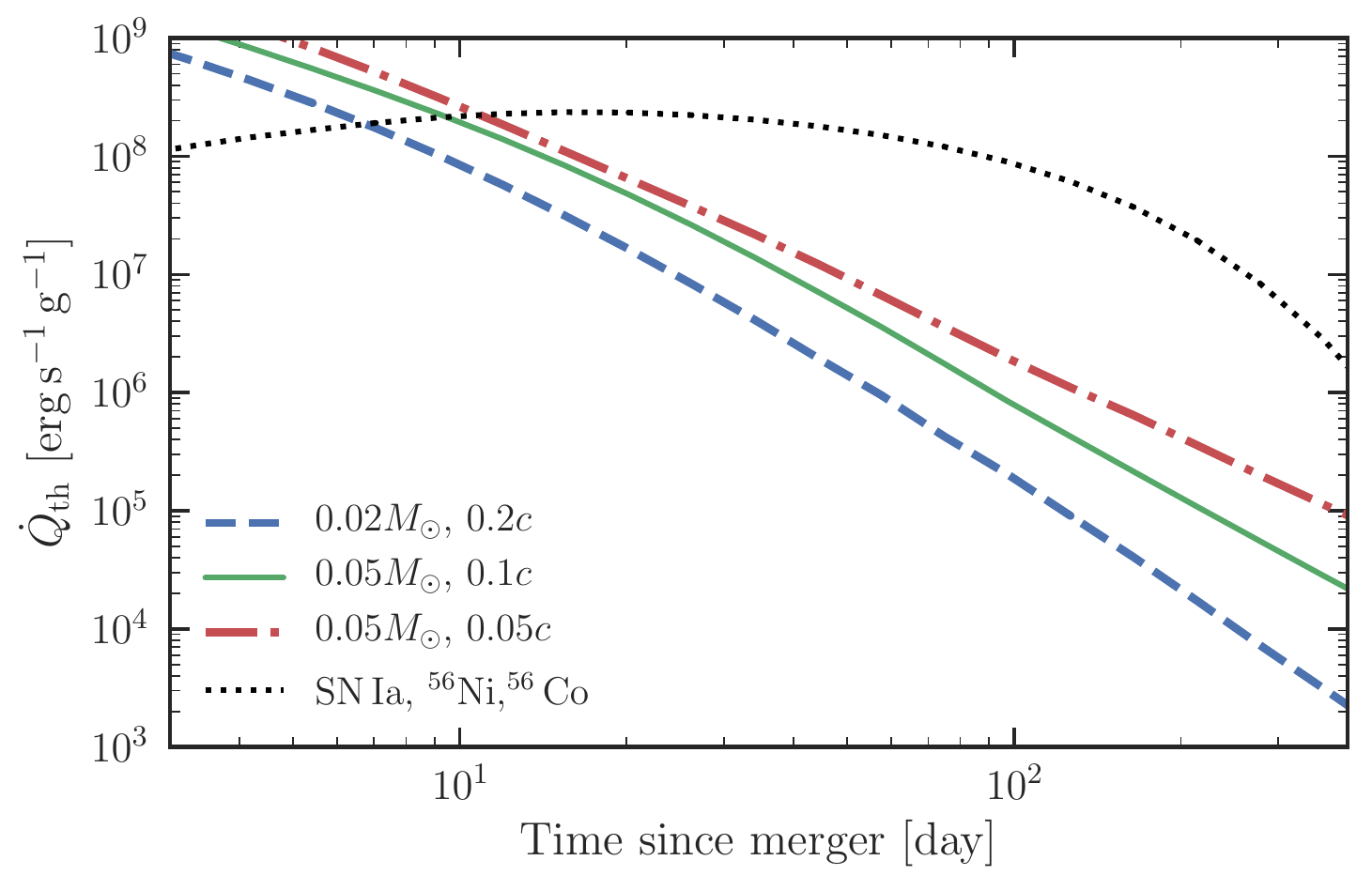}
\includegraphics[scale=0.5]{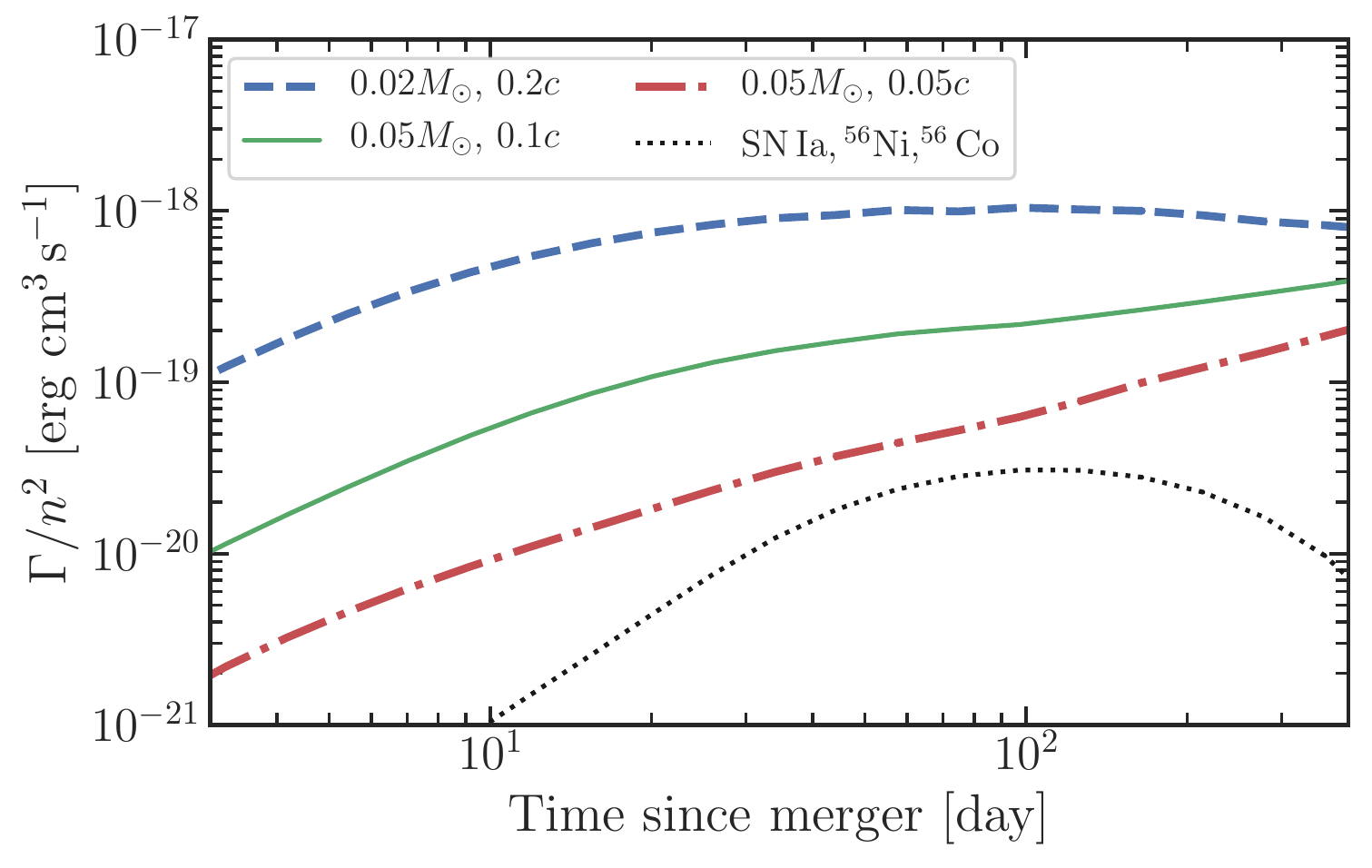}
\caption{Specific heating rates and
normalized heating rates of $\beta$-decay of $r$-process nuclei. The final composition is assumed to be the solar $r$-process abundance pattern in a mass range of $130\leq A\leq 209$. Also shown for comparison is the specific heating rate of the decay chain $^{56}$Ni$\rightarrow^{56}$Co$\rightarrow ^{56}$Fe denoted as SN Ia. For calculating the thermalization in SNe Ia efficiencies, we assume $M_{\rm ^{56}Ni}=0.54M_{\odot}$, $M_{\rm ej}=1.4M_{\odot}$, and $v_0=7000\,{\rm km\,s^{-1}}$. Here the heating rates are computed by using an open code \citep{hotokezaka2020ApJ}.}\label{fig:heating}
\end{figure}

In the case that the ejecta material is mainly composed of $r$-process elements,
$\kappa$ is typically $\approx 10\,{\rm cm^2g^{-1}}$ for lanthanide-rich material and $\approx 1\,{\rm cm^2g^{-1}}$ for lanthanide-free material \citep{barnes2013ApJ,kasen2013ApJ,tanaka2013ApJ,Wollaeger,tanaka2020}. 
 The time when a kilonova enters the NSM nebular phase is roughly estimated by $\tau(t_{\rm thin})\approx 1$:
 \begin{eqnarray}
t_{\rm thin} & \approx & \sqrt{\frac{\kappa M_{\rm ej}}{4\pi v_{0}^2}} ,\\
& \approx & 35\,{\rm day}\,
\left(\frac{\kappa}{10\,{\rm cm^2g^{-1}}} \right)^{1/2} 
 \left(\frac{M_{\rm ej}}{0.05M_{\odot}} \right)^{1/2} 
 \left(\frac{v_{0}}{0.1c} \right)^{-1},\label{eq:thin}
\end{eqnarray}
where $c$ is the speed of light.
Note that this time scale is significantly shorter for lanthanide-free material and fast expanding material. For instance, $t_{\rm thin}$ is $\approx 2\,{\rm day}$ for a lanthanide-free ejecta with $M_{\rm ej}=0.02M_{\odot}$, $\kappa=0.5\,{\rm cm^2 g^{-1}}$, and $v_0=0.25c$. 
 
 We assume the radioactivity of $r$-process nuclei as the source of heat and ionization. Note that the light curve of the  GW170817 kilonova is consistent with the picture that $\beta$-decay of $r$-process nuclei predominantly  heats the ejecta material 
over the time scales from $0.5$ to 70 day (e.g., \citealt{Kasliwal2019MNRAS})\footnote{Spontaneous fission and $\alpha$-decay of heavy nuclei can also be the energy source \citep{Zhu2018ApJ,Wanajo2018ApJ,Wu2019PhRvL}}. In the nebular phase,
the ejecta is  optically thin for 
$\gamma$-rays, and hence,
we consider only $\beta$-decay electrons.  We use the heating rate  with $130\leq A \leq 209$, where $A$ is atomic mass number, provided by \cite{hotokezaka2020ApJ}.
In the following, we describe the  characteristic features of the $\beta$-decay heating rate relevant to the NSM nebular modelings.

At early times,  the  heating rate per unit mass approximately follows  \citep{metzger2010MNRAS}:
\begin{eqnarray}
\dot{Q}_{\rm th}(t) \propto t^{-1.3} \label{eq:heat}
\end{eqnarray}
This power law is valid as long as  thermalization of $\beta$-decay electrons occurs on a time scale much shorter than a dynamical time. The heating rate starts to deviate from equation (\ref{eq:heat}) around the thermalization time, $t_{\rm th}$, estimated by  
$\tau_{\rm eff,e}=\kappa_{\rm eff,e}\rho c t\sim 1$:
\begin{eqnarray}
t_{\rm th} &\approx & \left(\frac{C_{\rho}c  \kappa_{e,{\rm eff}} M_{\rm ej}}{ v_0^3} \right)^{1/2},\nonumber \\
& \approx & 55\,{\rm day}\,\left(\frac{C_{\rho}}{0.05}\right)^{1/2}
\left(\frac{M_{\rm ej}}{0.05M_{\odot}}\right)^{1/2} \left(\frac{v_{0}}{0.1c}\right)^{-3/2} \nonumber  \\
& & \times \left(\frac{\kappa_{{\rm eff},e}}{4.5\,{\rm cm^2g^{-1}}}\right)^{1/2}
 \left(\frac{E_{e}}{0.25\,{\rm MeV}}\right)^{-1/2},\label{eq:thermal}
\end{eqnarray}
where  $E_e$ is the initial energy of $\beta$-decay electrons and $\kappa_{{\rm eff},e}$ is an effective opacity of the interaction of $\beta$-decay electrons with the ejecta material. Hereafter, we use the one-zone approximation, in which the density at a given time is represented by the mass weighted mean, $\rho_m(t)=C_{\rho} M_{\rm ej}t^{-3}v_0^{-3}$, where $C_{\rho}$ is a normalization constant \citep{hotokezaka2020ApJ}.

For $t\gtrsim t_{\rm th}$, the specific heating rate declines as \citep{Kasen2019ApJ,waxman2019ApJ,hotokezaka2020ApJ}
\begin{eqnarray}
\dot{Q}_{\rm th}(t)\propto t^{-2.8}~~~({\rm for}~t\gg t_{\rm th}).\label{eq:th}
\end{eqnarray}
This break in the $\beta$-decay heating rate from $\propto t^{-1.3}$ to $\propto t^{-2.8}$ is referred to as the thermalization break, which typically occurs in the nebular phase (see equations \ref{eq:thin} and \ref{eq:thermal}).

It is useful to introduce the normalized heating rate:
\begin{eqnarray}
\frac{\Gamma}{n^2} & = & \frac{\rho_m\dot{Q}_{\rm th}}{n^2}
 \propto   
\left\{ \begin{array}{rl}
 t^{1.7}
 &\mbox{   ($t\ll t_{\rm th}$)}, \\
 t^{0.2}
 &\mbox{    ($t\gg t_{\rm th}$)}.
       \end{array} \right.\label{eq:ng}
\end{eqnarray}
where $\Gamma$ is the heating rate per unit volume and
$n$ is the mass-weighted mean atomic number density
\begin{eqnarray}
n 
& \approx & 4\cdot 10^{4}\,{\rm cm^{-3}}\,
\left(\frac{\langle A \rangle}{150}\right)^{-1}
\left(\frac{C_{\rho}}{0.05}\right)\nonumber  \\
& & \times \left(\frac{v_{0}}{0.1c}\right)^{-3} \left(\frac{M_{\rm ej}}{0.05M_{\odot}}\right)
\left(\frac{t}{35\,{\rm d}}\right)^{-3},
\end{eqnarray}
where   $\langle A \rangle$ is the  mean atomic mass of the ejecta material. 
As we will see later, the evolution of kinetic temperature and ionization degree roughly follows the evolution of the normalized heating rate.

Figure \ref{fig:heating} shows the specific heating rates, $\dot{Q}_{\rm th}$, and normalized heating rates, $\Gamma/n^2$, with three different combinations of the ejecta mass and velocity. For more massive and slower ejecta, the  normalized heating rate at a given time is smaller, corresponding to that the efficiency of ionization and heating is lower. As expected from equation (\ref{eq:ng}), the slope of the normalized heating rates becomes almost flat at later times. For comparison, figure \ref{fig:heating} also shows the heating rate of the decay chain powering SNe Ia, $^{56}$Ni$\rightarrow ^{56}$Co$\rightarrow^{56}$Fe, with $M_{\rm ^{56}Ni}=0.54M_{\odot}$, $M_{\rm ej}=1.4M_{\odot}$ and $v_0=7000\,{\rm km\,s^{-1}}$. Unlike the $r$-process cases the normalized heating rate of this decay chain turns to decrease around the half-life of $^{56}$Co. Note that the normalized heating rate of NSMs is much larger than that of SNe Ia because of the difference in the expansion velocity, suggesting that ionization in the 
NSM ejecta is more efficient.

 
The ionization rate of an $i$-th ionized ion, $X^{i+}$, by $\beta$-decay electrons is characterized by the work per ion pair $w_i$ (see Appendix \ref{sec:w}). 
With this quantity, the ionization rate per unit volume  is  given by
\begin{eqnarray}
\Upsilon_i =\frac{\Gamma}{w_i}.
\end{eqnarray}
For the NSM nebulae, we estimate $w_i/I_{i,1}\sim 30$, where 
$I_{i,1}$ is the first ionization potential of $X^{i+}$. This value
indicates that the significant fraction of $\beta$-electrons' energy is deposited to the thermal energy and 
only $\sim 3\,\%$ of it is consumed by ionization. Therefore, we neglect the recombination continuum cooling.


\section{Equations for nebula modeling} \label{sec:eqs}
In the nebular phase, the ejecta material is in non-LTE, i.e., only free electrons are distributed according to Maxwell' law and atoms are not in equilibrium. Thus, 
one must solve the ionization and thermal balance to obtain the kinetic temperature, $T_e$, and ionization fractions.
Here we use the nebular modeling 
developed by \cite{Axelrod1980} with some modifications. 
In this section, we briefly describe the equations used  and discuss some generic features of the NSM nebular emission that arise from the characteristic properties of the $r$-process heating rate (equation \ref{eq:ng}) without specifying the details of the atomic structure. 

We consider the NSM nebular phase where the  recombination  and cooling time scales are  shorter than a dynamical time. This condition allows us  to use the steady state approximation. As we will show later, it holds  $t\lesssim 100\,{\rm day}$ after merger.
We assume  that the ejecta is composed of
a single atomic species, $X$, for simplicity\footnote{We consider a single atomic species only for ionization and thermal balances. However we consider that the $\beta$-decay heat is produced by many different isotopes.}.
Under these conditions, 
the equation for ionization balance is 
\begin{eqnarray}
-\Upsilon_i f_i & - &\sum_{j>i}P_{ij}\alpha_{j+1}\chi  f_{j+1}n^2 +  \left(1-P_{ii} \right)\alpha_{i+1}\chi  f_{i+1}n^2\approx 0\nonumber \\
& & ~~~~~~~~~~~~~~~~~~~~~~{\rm for}~0\leq i \leq N-1,\label{eq:i1}
\end{eqnarray}
where 
$f_i$ is the number fraction of $X^{i+}$, $P_{ij}$ is the probability that the photons created by the recombination of an ion $X^{j+}$ ionize an ion $X^{i+}$, $\alpha_i$ is the recombination rate coefficient for $X^{(i+1)+}\rightarrow X^{i+}$, and $\chi$ is the free electron fraction. 
This equation can be rewritten in the form \citep{Axelrod1980}
\begin{eqnarray}
-\frac{\Gamma}{n^2 } \frac{f_i}{w_i} & - & \sum_{j>i}P_{ij}\alpha_{j+1}\chi  f_{j+1} +  \left(1-P_{ii} \right)\alpha_{i+1}\chi  f_{i+1}\approx 0\nonumber \\ & &
~~~~~~~~~~~~~~~~~~~{\rm for}~0\leq i \leq N-1,\label{eq:i2}
\end{eqnarray}
The ion fractions $f_i$ and free electron fraction $\chi$ are obtained by solving equation (\ref{eq:i2}) together with the conditions of $\sum _{i=0}^Z f_i=1$ and $\chi = \sum_{i=1}^Z i f_i$ for a given $n$ and $T_e$. The details of the reprocess of recombination radiation, $P_{ij}$, are described in Appendix \ref{app:photo}.

\begin{figure*}
\begin{center}
\includegraphics[scale=0.35]{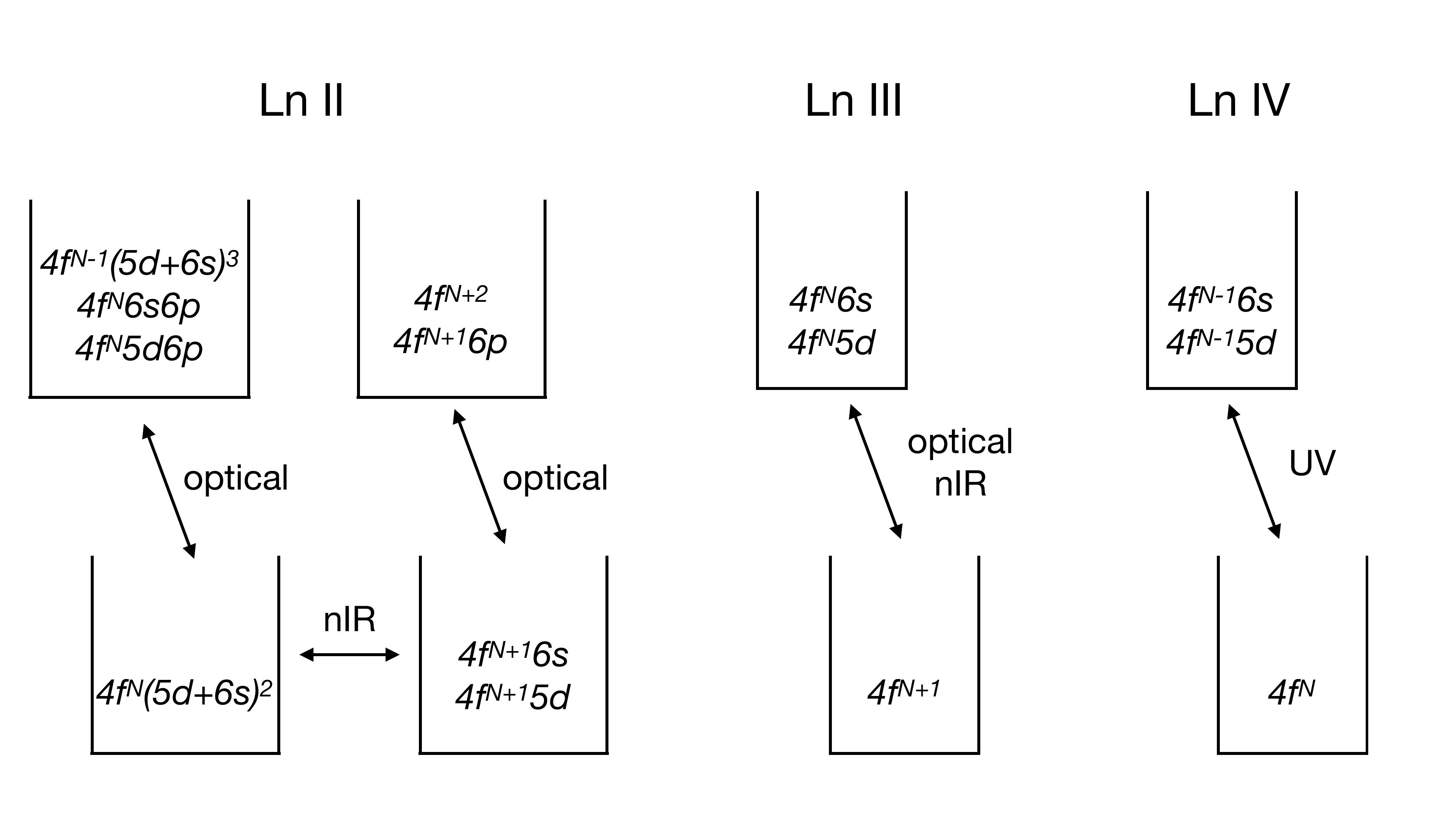}
\end{center}
\caption{
Characteristic structure of the transitions from the lowest configurations for second to fourth lanthanide ions. {\it Ln II}: The ground configuration of Ln II is one of the configurations in the two groups $4f^N(5d+6s)^2=4f^N5d^2,\,4f^N5d6s,\,4f^N6s^2$ and  $4f^{N+1}5d,\,6s$. The energy levels of these configurations overlap and there exist permitted transitions between the two groups. This feature causes  dense spectra in the nIR band. The transition lines from  these lowest energy groups to the second lowest ones are typically in the optical region.
{\it Ln III and IV}: The ground configurations of Ln III and Ln IV are $4f^{N+1}$ and $4f^{N}$, respectively, except for La III, Gd III, and Lu III. 
The typical energy separations  between the ground and the next lowest configurations are in the nIR to optical region for Ln III and in the ultraviolet region for Ln IV. Because only excited configurations overlap  their spectra are less dense than those of Ln II. 
}
\label{fig:configuration}
\end{figure*}

The thermal balance determines the temperature $T_e$ and level population:
\begin{eqnarray}
\Gamma - \sum_{i}\Lambda_i\approx 0,\label{eq:t1}
\end{eqnarray}
where  $\Lambda_i$ is the cooling rate of $X^{i+}$   per unit volume.  This equation can be rewritten as
\begin{eqnarray}
\frac{\Gamma}{n^2} & - & \chi \sum_{i}\left(\frac{\Lambda_i}{n_in_e}\right)f_i  \approx  0,\label{eq:t2}
\end{eqnarray}
where $n_i=f_in$ and $n_e = \chi n$.
Note that the cooling rate due to free-free emission is much smaller than the atomic cooling rate in the temperature range of the nebular phase and therefore we consider only the atomic cooling in the following. 

The number density of $X^{i+}$ in a level $j$, $n_{i,j}$, is determined by a given  kinetic temperature, density, electron fraction, and radiation field.
We use the escape probability approximation to solve the level population (e.g., Chapter 19 of \citealt{Draine}):
\begin{eqnarray}
\frac{dn_i}{dt} = \sum_{j<i}\left[n_jn_ek_{ji}-n_in_ek_{ij}-\langle \beta_{ij}\rangle n_iA_{ij}\right] \nonumber \\
+\sum_{j>i}\left[n_jn_ek_{ji}-n_in_ek_{ij}+\langle \beta_{ij}\rangle n_jA_{ji}\right]\approx 0\label{eq:level},
\end{eqnarray}
where  $k_{ij}$ is the collisional rate coefficient for $i\rightarrow j$, $A_{ij}$ is the
radiative transition rate for $i\rightarrow j$, and $\langle \beta_{ij}\rangle$ is the escape probability of photons created by the transition $i\rightarrow j$.
Note that we omitted the suffix that denotes the ionizing state in equation (\ref{eq:level}).
Here we use the Sobolev optical depth to evaluate $\langle \beta_{ij}\rangle$ (see Appendix \ref{app:esc}). Because this description  includes only self-absorption of lines, the cooling function and spectrum of each ion can be computed separately. However, this approximation is not valid around the frequencies where the effect of the line overlapping is important. 
Such a situation can occur in the optical region for lanthanide-rich nebulae as will be discussed in \S \ref{sec:atom}.


The atomic cooling rate of $X^{i+}$ is calculated by
\begin{eqnarray}
\Lambda_i = \sum_{j>0}n_{i,j}\sum_{k<j}E_{jk}\langle \beta_{jk}\rangle A_{jk},
\end{eqnarray}
where   $E_{jk}$ is the energy difference between levels $j$ and $k$.  One can show that the normalized cooling rate $(\Lambda_i/n_en_i)$ depends only on $T_e$ at sufficiently low densities.
For NSM nebulae, as we will show later, $(\Lambda_i/n_en_i)$ is almost independent of the density at $n\lesssim 10^4\,{\rm cm^{-3}}$, corresponding to $t\gtrsim 40\,{\rm day}$ for $M_{\rm ej}=0.05M_{\odot}$ and $v_0=0.1c$.

The ionization degree, $f_i$, free  electron fraction, $\chi$, and kinetic temperature, $T_e$, at each time are obtained  by solving  equations (\ref{eq:i2}) and (\ref{eq:t2}) iteratively for given $\Gamma(t)$ and $n(t)$. Roughly speaking, equations (\ref{eq:i2}) and (\ref{eq:t2}) explicitly depend on time only through $\Gamma/n^2$. Therefore the thermodynamic quantities evolve with time according to $\Gamma/n^2$. This fact and equation (\ref{eq:ng})  suggest that the ionization degree,  electron fraction, and kinetic temperature  roughly increase as $\propto t^{1.7}$ for $t\lesssim t_{\rm th}$ and increase very slowly as $\propto t^{0.2}$ for $t\gtrsim t_{\rm th}$ (figure \ref{fig:heating}). This property is somewhat naturally expected from the fact that almost all the processes relevant to the ionization and thermal balance after the thermalization breaks are two-body collision between electrons and ions.

\section{Atomic properties of Neodymium}\label{sec:atom}
NSM ejecta are  composed of atoms with a wide range of atomic numbers, $Z\gtrsim 30$, in reality. The  experimental atomic data of these heavy elements  are largely unavailable.  To derive the atomic data necessary for our purpose we use atomic structure codes, General Relativistic Atomic Structure Package (\texttt{GRASP2K}; \citealt{grasp2013}) and Hebrew University Lawrence Livermore Atomic (\texttt{HULLAC}; \citealt{hullac2001JQSRT}) codes. 
\texttt{HULLAC} is an integrated code for calculating atomic structures and cross sections for the modelings of atomic processes in plasmas and emission spectra, which employs a parametric potential method for calculations of bound- and free-electron wavefunctions. The \texttt{GRASP2K} code provides more rigorous bound-electron wavefunctions based on the multiconfiguration Dirac–Hartree–Fock  method, which enables more ab-initio calculations of atomic structures and bound-bound radiative transition probabilities, and therefore,  
 we use \texttt{GRASP2K} to derive the level spectra and radiative transition rates (see \citealt{Gaigalas2019ApJS} for details) and compute recombination rate coefficients
by using \texttt{HULLAC}. 


\begin{figure*}
\begin{center}
\includegraphics[scale=0.5]{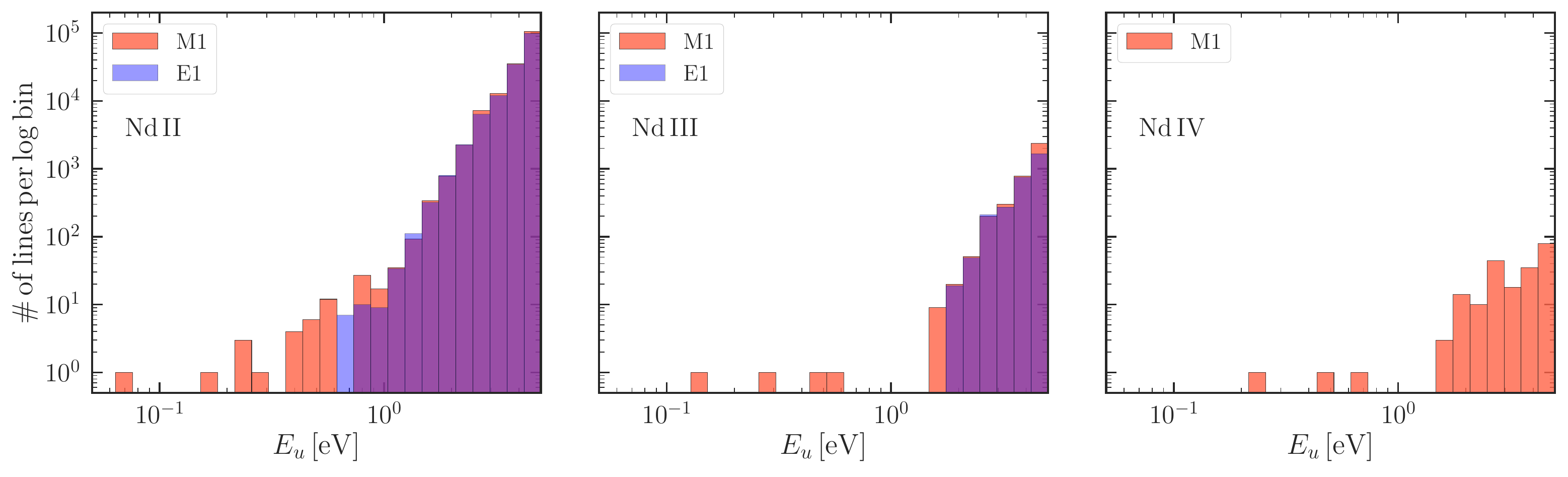}\\
\includegraphics[scale=0.5]{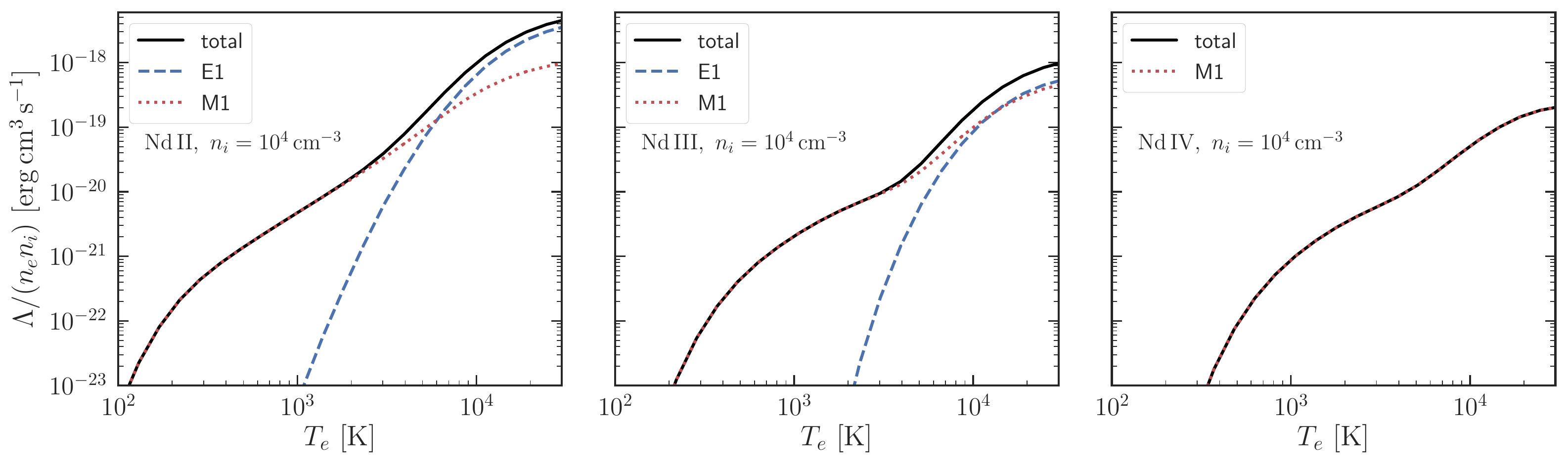}
\caption{
Number of transition lines per logarithmic interval of energy of excited states $E_u$ computed by using \texttt{GRASP2K} ({\it top panels}). Red and blue histograms show M1 and E1 transitions, respectively. Cooling function of Nd II, Nd III, and Nd IV at an atomic number density of $10^{4}\,{\rm cm^{-3}}$ ({\it bottom panels}). This number density corresponds to $\sim 40$ day after merger in the case of an ejecta mass of $\sim 0.05M_{\odot}$ and  expansion velocity of $\sim 0.1c$. Dotted and dashed curves show the cooling due to M1 and E1 transitions, respectively.
For Nd II and Nd III, the contribution of E1 transitions is significant at $T_e \gtrsim 5000\,{\rm K}$. 
}
\label{fig:A}
\end{center}
\end{figure*}

Lanthanide ions  enhance the opacity of NSM ejecta \citep{kasen2013ApJ,tanaka2013ApJ,tanaka2020,Barnes2020}, and thus, they are naturally expected to be strong emitters in the nebular phase. 
In addition, one may be able to capture, at least qualitatively, some important features of the nebular emission of lanthanide-rich ejecta by using a single element because of the similarity in the spectral structure between lanthanide elements. Motivated by these,
 we focus on  neodymium (Nd),  in order to qualitatively understand the nebular emission of lanthanide-rich ejecta in this and following sections.

\subsection{Characteristic structure of lanthanide ions}
Before proceeding the details of the atomic data, here, we briefly summarize some spectral properties of lanthanide ions \citep{GOLDSCHMIDT1978}. Lanthanide elements, Ln, are a group of elements with atomic numbers $57$ -- $71$ (La -- Lu). Their ions are characterized by the number of electrons in $4f$-shell, $N$ or $N+1$, where we use a number $N=Z-57$, e.g., $N=0$ for La and $N=14$ for Lu.
Their  configurations lying at low energies often have one to three electrons in the outer shells, $5d$ and $6s$, which means that the energy scales of $4f$, $5d$, and $6s$ are similar so that several different configurations with the same parity consists of a group.  Figure \ref{fig:configuration} shows a characteristic spectral structure of  first to third  lanthanide ions (Ln II-IV). For two groups connected by arrows, there are permitted transitions between them.
Forbidden transitions between different orbital angular momenta, $L$, as well as those associated with the fine structure are also important for the cooling rate and the emergent spectra. The characteristic spectra of lanthanides are summarized as follows.
\begin{enumerate}
    \item Permitted (E1) transitions of Ln II and Ln III  exist in the nIR and optical bands. These lines lead to the enhancement of  absorption and can also be the source of nIR-optical emission in the nebular phase.
    \item Dipole forbidden (M1) transitions of Ln II - Ln IV between different configurations or between different total orbital angular momenta   produce emission lines in the nIR and optical bands. 
    \item  Transitions between  fine stricture levels produce mid-IR  lines ($\lambda \sim 10\,{\rm \mu m}$).
\end{enumerate}

 Note that,  among Ln II, Nd II has resonance lines at the lowest transition energy and more E1 transitions at longer wavelengths. In fact, \cite{tanaka2020} show that the abundance of Nd atoms has the most significant effect on the opacity in kilonovae. In the following, we focus on the atomic data of Nd.





\begin{figure}
\begin{center}
\includegraphics[scale=0.35]{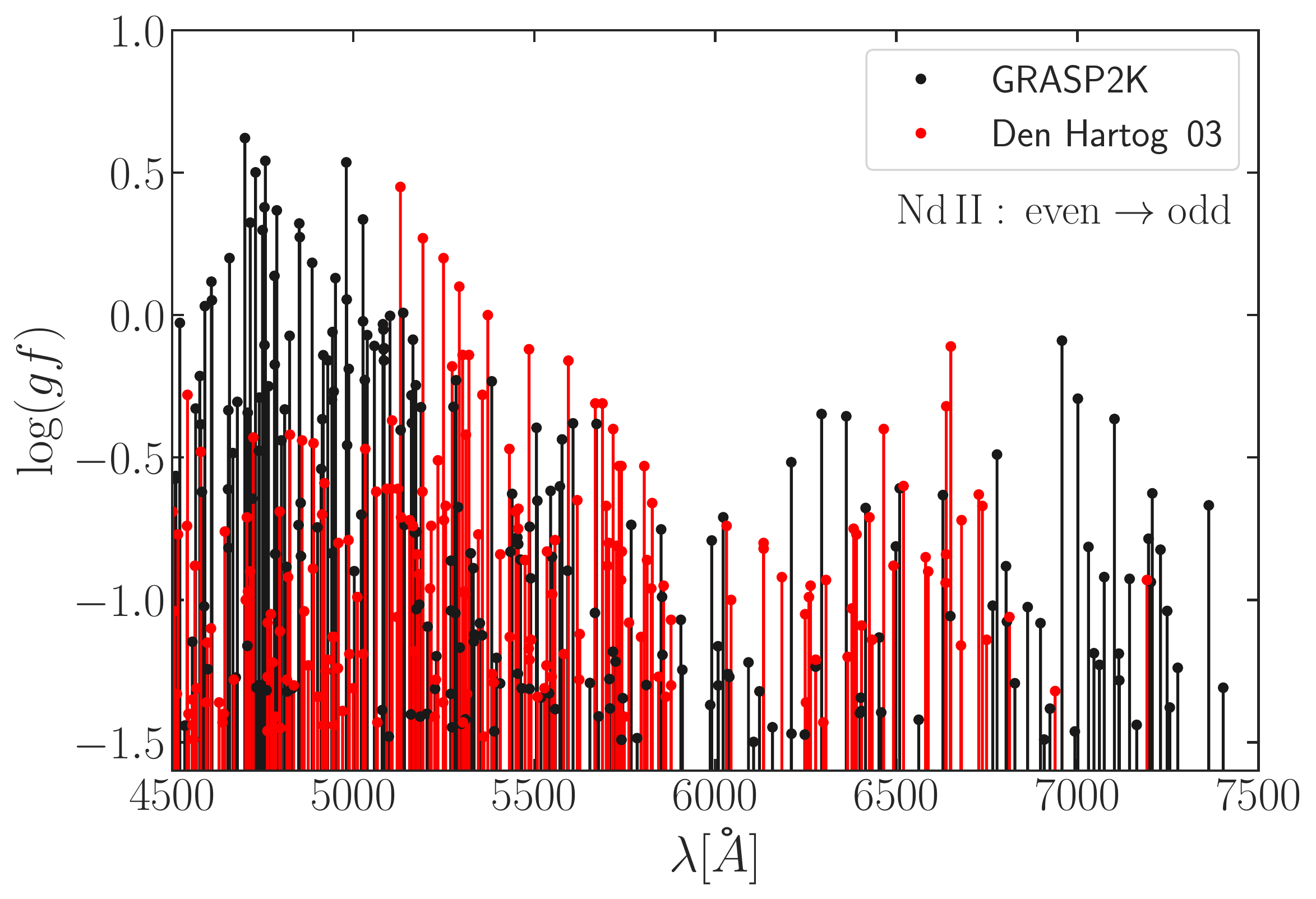}
\includegraphics[scale=0.35]{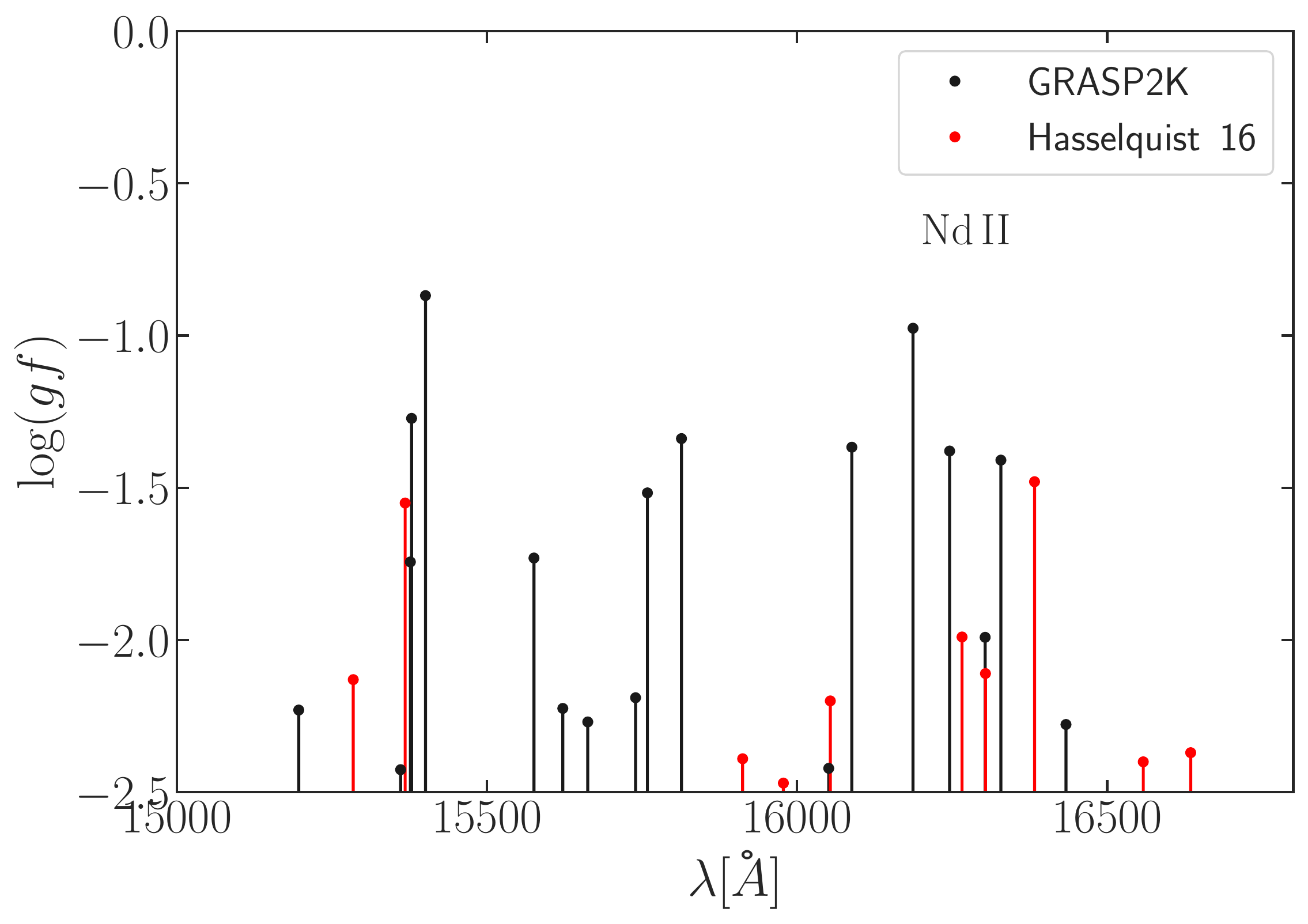}
\includegraphics[scale=0.35]{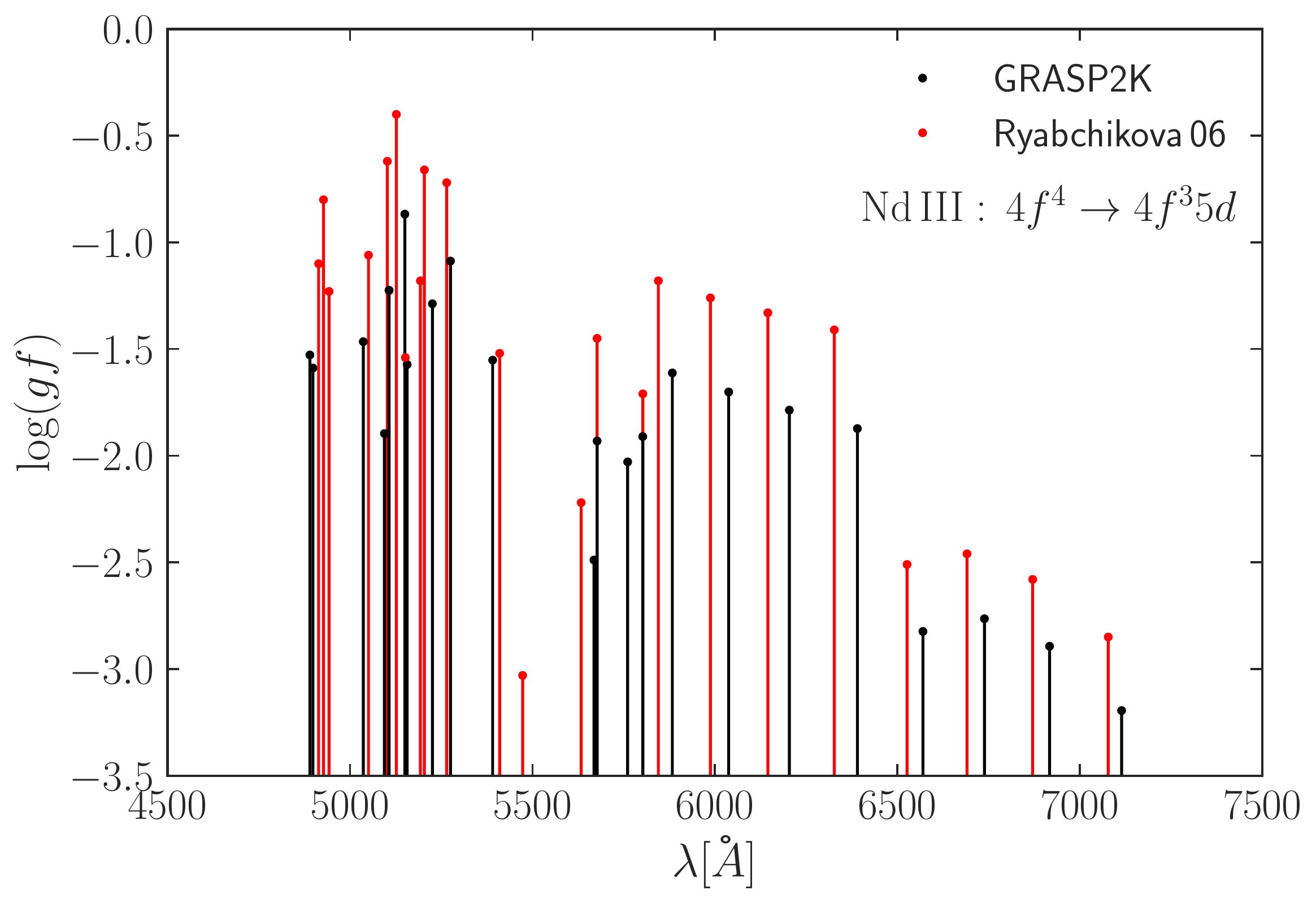}
\end{center}
\caption{
Comparison of line strengths of Nd II and Nd III computed by using \texttt{GRASP2K} \citep{Gaigalas2019ApJS} with  the experimental results \citep{Den2003} and the APOGEE line list \citep{Hasselquist2016ApJ} for Nd II and the Nd III line list based on the stellar spectrum of a strongly magnetic Ap star \citep{Ryabchikova2006}.
}\label{fig:lines}
\end{figure}

\subsection{Radiative transition rate}
We include the excited levels of Nd ions up to  $\approx 5$ eV \citep{Gaigalas2019ApJS}. The numbers of levels included are 1400, 200, 40 for Nd II, Nd III, and Nd IV, respectively. 
Figure \ref{fig:A} shows the distribution of E1 and M1 transitions.  Nd II has more lines than Nd III and Nd IV, suggesting that the cooling of Nd II per ion is the most efficient.  Note that radiative transition rates of M1 transitions are lower by a factor of $\sim 10^5$ than E1 transitions. 
We also examined E2  transitions and found that their contribution to the cooling function is rather minor in the relevant temperature range and therefore we decide not to include E2 transitions. 

Note that there are excited states that can decay through E1 transitions down to $\sim 0.7$\,eV for Nd II, indicating that the cooling through the E1 lines is important even around $5000\,{\rm K}$.  This feature  is qualitatively different from SN Ia nebulae, where the cooling is completely dominated by forbidden lines of the iron group elements. 

Figure \ref{fig:lines} compares the line spectra of Nd II and Nd III computed by \texttt{GRASP2K} with those from \cite{Den2003} and \cite{Ryabchikova2006} in the optical region and that from \cite{Hasselquist2016ApJ} in the nIR region.   
\cite{Den2003} experimentally measured the wavelengths and oscillator strengths of over 700 lines of Nd II. Here we focus on the intensive lines with $\log_{10} gf>-1.5$, $E_u<35000\,{\rm cm^{-1}}$, and $J\geq 5/2$ in the range of $4500\AA<\lambda < 7500 \AA$. The number of lines satisfying these restrictions is $\sim 180$. The line distribution of \texttt{GRASP2K} statistically agrees with the laboratory-based one. The \texttt{GRASP2K} line distribution is also roughly in agreement with the nIR lines of Nd II 
identified from the Apache Point Observatory Galactic Evolution Experiment (APOGEE) H-band spectra \citep{Hasselquist2016ApJ}. 

Because the line spectrum of Nd III is poorly known  experimentally, here we compare the \texttt{GRASP2K} result with the line list provided by  \cite{Ryabchikova2006}, in which they propose the line classification for Nd III based on stellar spectra and a theoretical calculation of atomic structure. In figure \ref{fig:lines}, we show 23 lines associated with the transitions between 4f$^4$ and 4f$^3$5d in the range of $4500\AA<\lambda < 7500 \AA$. We note that the wavelength of each line agrees within $\sim \%$ level. 
We consider that the \texttt{GRASP2K} line list  is sufficiently accurate at least for E1 transitions to capture the spectral structure of the NSM nebular emission.

\subsection{Collisional rate coefficient and critical density}\label{sec:coll}

We derive the collisional rate coefficients for the \texttt{GRASP2K} atomic data with the procedure described in Appendix \ref{app:coll}. 
Here we discuss the typical critical densities for Nd ions and implications to the evolution of the cooling functions and spectra.
The critical density for a given upper level $u$ is estimated as
\begin{eqnarray}
n_{{\rm crit},u} & \equiv & \frac{\sum_{l<u}A_{ul}}{\sum_{l<u}k_{ul}},\\
&\sim &
\left\{ \begin{array}{rl}
 10^9\,{\rm cm^{-3}} 
 &\mbox{   (E1 transition)}, \\
 10^4\,{\rm cm^{-3}}
 &\mbox{     (M1 transition)},
       \end{array} \right.
\end{eqnarray}
where we used the typical value of $k_{ul}$ and $A_{ul}$.
These critical densities correspond to the critical times:
\begin{eqnarray}
t_{{\rm crit}} & \sim & 
\left\{ \begin{array}{rl}
 1\,{\rm day} 
 &\mbox{   (E1 transition)}, \\
 40\,{\rm day}
 &\mbox{     (M1  transition)}.
       \end{array} \right.
\end{eqnarray}
{When the  NSM ejecta becomes optically thin,
the time scale of E1 radiative deexcitation is much faster than that of excitation, i.e.,  
$t_{\rm crit}{\rm (E1)}\ll t_{\rm thin}$. Therefore, the level populations in the nebular phase are always far from those in collisional equilibrium, i.e., the LTE values.  }
For $t>t_{\rm crit}{\rm (M1)}$, excited levels predominantly decay through radiative transition. Such a state is referred to as corona equilibrium.
In this case,  the cooling rate is  proportional to $n_en_i$, i.e., the cooling function, $\Lambda_i/n_en_i$, is independent of the density, and therefore, 
the kinetic temperature is expected  to evolve very slowly with time after $t_{\rm crit}{\rm (M1)}$ because of $\Gamma/n^2 \propto t^{0.2}$.

\subsection{Cooling function}


The bottom panels of figure \ref{fig:A} show the cooling functions of  Nd II, III and IV ions at an ion density of $10^{4}\,{\rm cm^{-3}}$, corresponding to $\sim 40\,$ days after merger for $M_{\rm ej}=0.05M_{\odot}$ and $v_0=0.1c$. We find the overall trend of the cooling functions, $\Lambda({\rm IV})< \Lambda({\rm III}) < \Lambda({\rm II})$, which
can be understood from the fact that Nd II has more lines in the IR to optical region. M1 transitions dominate the cooling functions of Nd II and Nd III for $T_e<6000$\,K and $<10000$\,K, respectively. This feature is expected from the characteristic  lanthanide spectra (figure \ref{fig:configuration}).

\begin{figure*}
\begin{center}
\includegraphics[scale=0.3]{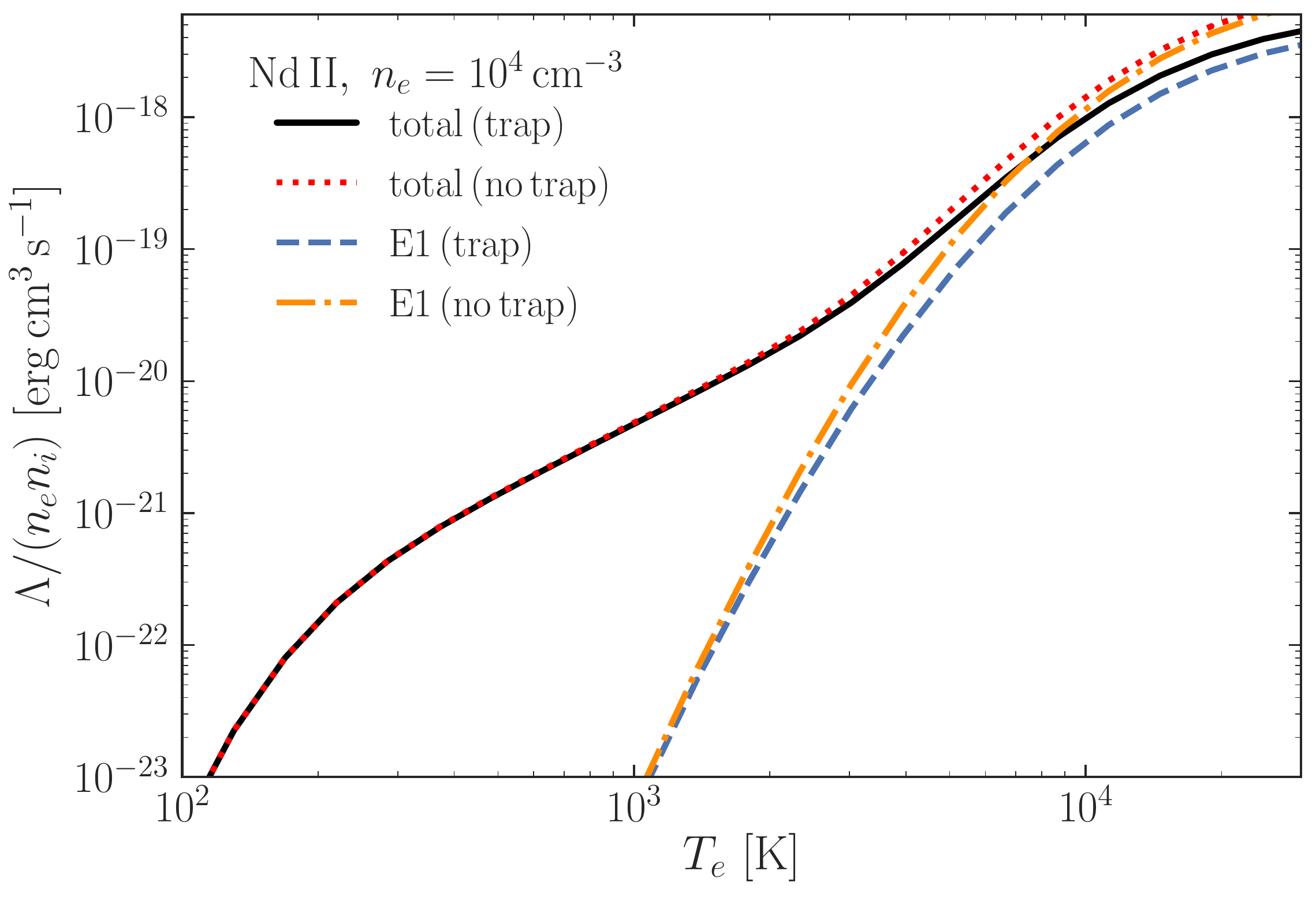}
\includegraphics[scale=0.3]{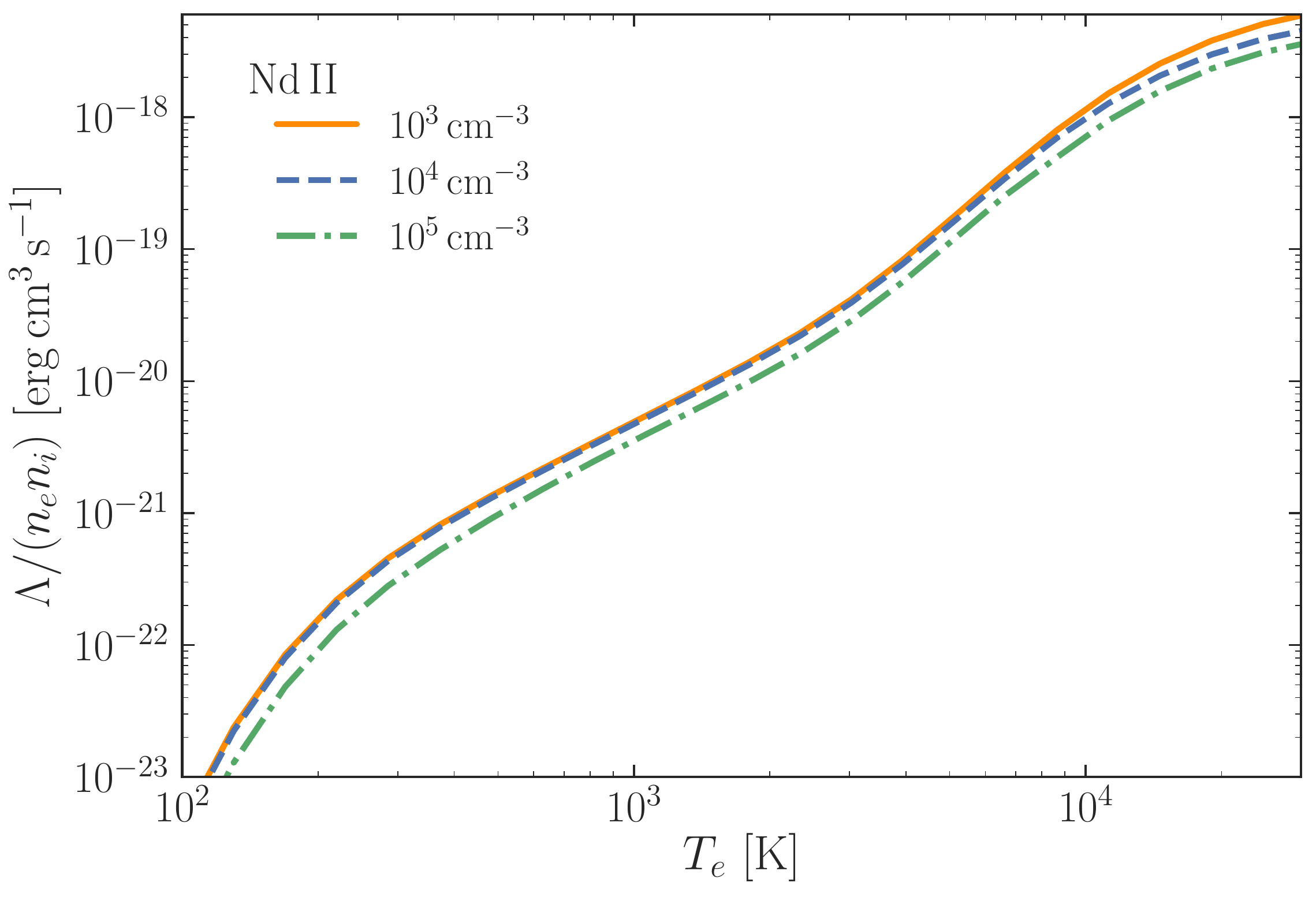}
\end{center}
\caption{
Cooling function of Nd II with and without the radiation trapping effect ({\it left}) and  at different densities ({\it right}).
}
\label{fig:cooling2}
\end{figure*}

Figure \ref{fig:cooling2}  depicts the effect of self-absorption ({\it left}) and the density effect ({\it right}) on the cooling rates. The cooling functions without the trapping effect are calculated with the assumption of $\langle \beta_{ij} \rangle=1$.
We note that the absorption effect reduces the cooling function of Nd II by $\gtrsim 50\%$ at $\gtrsim 10^4$\,K, where E1 transitions dominate the cooling rate. This effect is weaker for Nd III and absent for Nd IV. Note that, however, we likely overestimate the escape probability of lines with $\lambda \lesssim 1\,{\rm \mu m}$  because these lines may be  absorbed by nearby permitted lines such as resonance lines (see more details in \S \ref{sec:evolve}).
The density effect is  quite small at the densities of lanthanide-rich NSM nebulae. Thus, for $n\lesssim 10^{4}\,{\rm cm^{-3}}$, the ejecta is in corona equilibrium and the cooling function can be considered  to be independent of the density. For the results presented in the following section, the trapping and density effects are accounted for.

The cooling time scale is estimated as
\begin{eqnarray}
t_{\rm cool}& \approx & \frac{kT_e}{\Lambda/n} \sim 10^{2}\,{\rm s}\,\left(\frac{\Lambda/n_en}{10^{-19}{\rm erg\,cm^3s^{-1}}} \right)^{-1}
\left(\frac{T_e}{10^4\,{K}}\right)\nonumber \\ 
& & \times \left(\frac{M_{\rm ej}}{0.03M_{\odot}} \right)^{-1}
\left(\frac{\langle A \rangle}{150} \right)
 \left(\frac{v_0}{0.1c} \right)^{3}
 \left(\frac{t}{30{\rm day}} \right)^{3},
\end{eqnarray}
where $\Lambda$ is the total cooling function.
This time scale is much shorter than a dynamical time until $\sim 10$ years after merger, and thus,  the steady-state approximation for thermal balance is valid on the time scale, $t\lesssim 100\,{\rm day}$, focused in this work.

\begin{figure*}
\begin{center}
\includegraphics[scale=0.55]{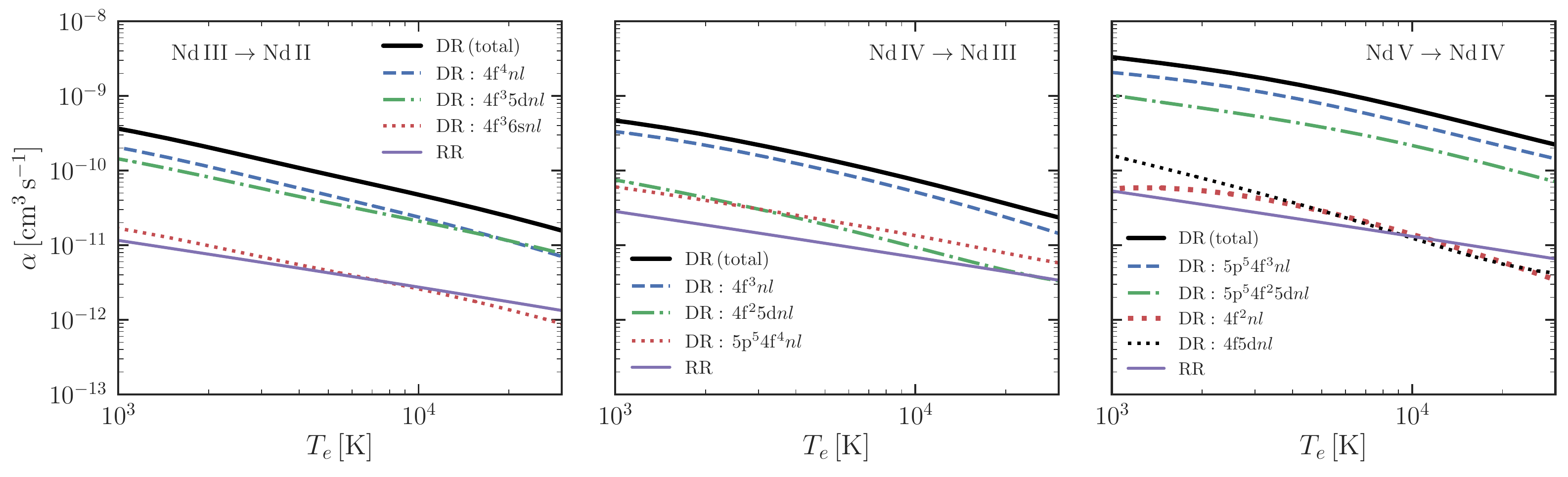}
\end{center}
\caption{
Rate coefficients for dielectronic recombination (DR) and radiative recombination (RR). The rate coefficients of dielectronic recombination are obtained by using \texttt{HULLAC}.
For dielectronic recombination, each line shows the contribution of a specific configuration of the autoionizing state $\gamma nl$, where $\gamma$ denotes the core configuration, $n$ and $l$ denote the principal and orbital angular momentum quantum numbers of the captured electron. The range of $n$ and $l$ of each autoionizing state included in the calculation are described in the text.
}
\label{fig:recombination}
\end{figure*}

\subsection{Dielectronic recombination}\label{sec:rec}
 Dielectronic recombination dominates over radiative recombination for lower ionized Nd ions. At nebular temperatures ($\sim 10^4\,{\rm K}$),  autoionizing states lying  slightly above  the ionization threshold contribute to the dielectronic capture process so that resolving fine structure is important here.  For this purpose, we use the level mode of \texttt{HULLAC} to obtain the energy levels, radiative transition rates, and autoionization rates. With these quantities, we  calculate the rate coefficients by following the prescription of \cite{Nussbaumer1983A&A} (see also Appendix \ref{app:rec}).
We include the following autoionizing states: 
\begin{itemize}
    \item Nd II: 4f$^4nl$ ($n\leq 10,\,l\leq 5$), 4f$^35$d$nl$ ($n\leq 8,\,l\leq 5$), and 4f$^36$s$nl$ ($n\leq 7,\,l\leq 4$)
    \item Nd III: 4f$^3nl$ ($n\leq 13,\,l\leq 5$), 4f$^25$d$nl$ ($n\leq 8,\,l\leq 5$), and 5p$^5$4f$^4nl$ ($n\leq 6,\,l\leq 3$)
    \item Nd IV: 4f$^2nl$ ($n\leq 11,\,l\leq 5$), 4f$^15$d$nl$ ($n\leq 7,\,l\leq 4$), 5p$^5$4f$^3nl$ ($n\leq 8,\,l\leq 3$), and 5p$^5$4f$^2$5d$nl$ ($n\leq 6,\,l\leq 3$)
\end{itemize}
Here an autoionizing state is denoted by $\gamma nl$, where $\gamma$ denotes the state of the core electrons, $n$ and $l$ denote the principal and orbital angular momentum quantum numbers of the captured electron. We note that the contribution of each configuration with higher $n$ and $l$ that is not included  is less than $\sim 1\%$ for $T_e \lesssim 10^4\,{\rm K}$.

Figure \ref{fig:recombination} shows the  recombination rate coefficients for Nd II - IV. The contribution of radiative recombination to the total rate coefficient is less than $10\%$ for all the cases. Note that, for Nd I, we assume that the rate coefficient of dielectronic recombination is $1/4$ of that of Nd II because of the limitation of computational time.

The recombination time scale is estimated as
\begin{eqnarray}
t_{\rm rec}  \sim  \frac{1}{\alpha n_e } & \sim & 5\,{\rm day}\,
\left(\frac{\alpha}{10^{-10}\,{\rm cm^3\,s^{-1}}} \right)^{-1}
\left(\frac{M_{\rm ej}}{0.05M_{\odot}} \right)^{-1}\nonumber \\
& & \times \left(\frac{\langle A \rangle}{150} \right)
 \left(\frac{v_0}{0.1c} \right)^{3}
 \left(\frac{t}{40\,{\rm day}} \right)^{3},
\end{eqnarray}
where $\alpha$ is the total recombination rate coefficient. 
The ionization time scale is estimated from the heating rate (see figure \ref{fig:heating}):
\begin{eqnarray}
t_{\rm ion}\sim \frac{1}{\Upsilon_{\rm Nd\,III}} \sim 5\,{\rm day}
 \left(\frac{t}{40\,{\rm day}} \right)^{2.8}~{\rm for}\,t\gtrsim t_{\rm th},
\end{eqnarray}
where we have used $w_i\sim 60\,$eV for Nd III. For the fiducial model, $M_{\rm ej}=0.05M_{\odot}$ and $v=0.1c$, these two time scales become comparable to a dynamical time at $\sim 100$ day. Thus, we consider the nebular phase at $\lesssim 100$ day after merger, where the steady-state approximation is valid.

\begin{figure*}
\begin{center}
\includegraphics[scale=0.29]{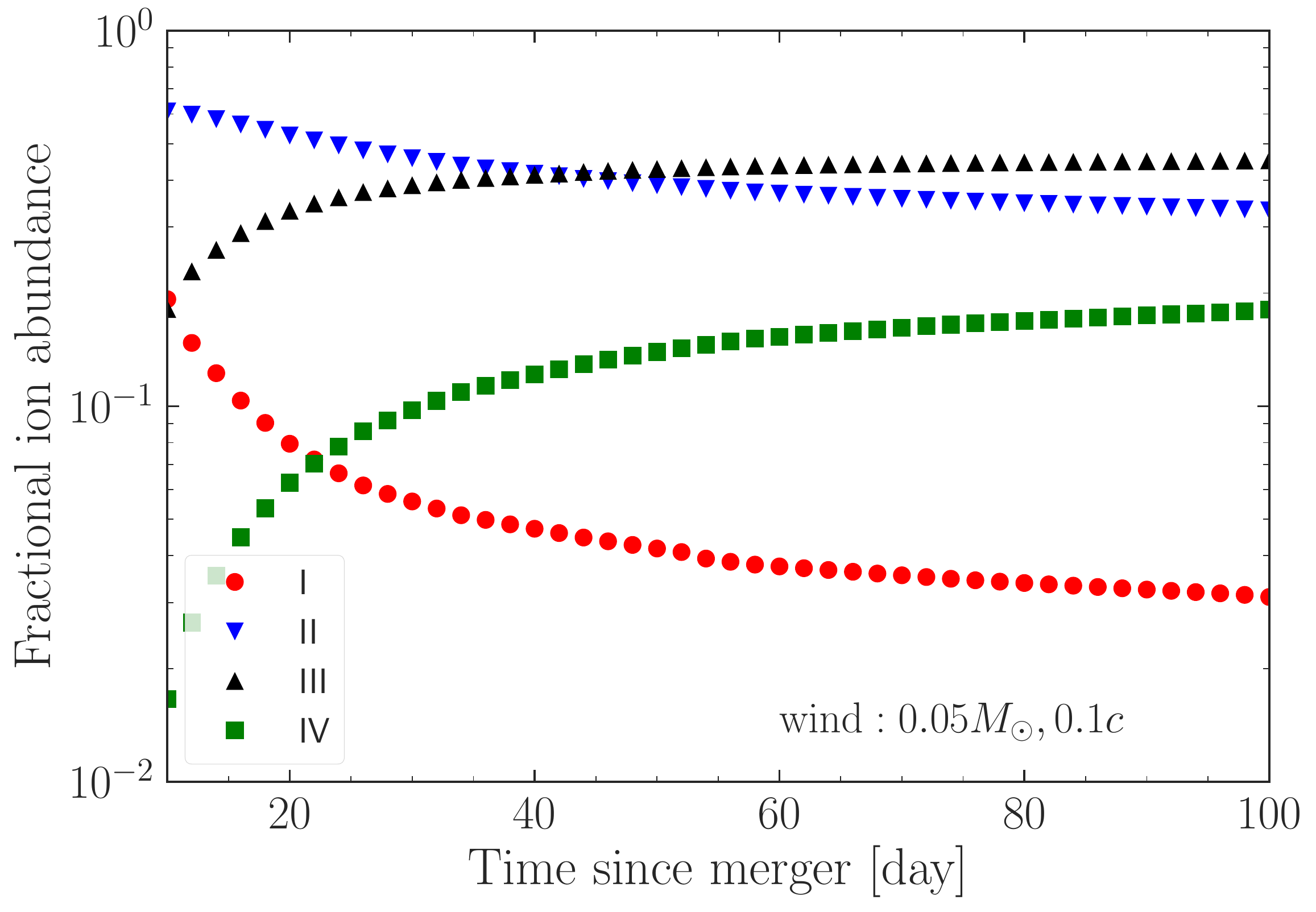}
\includegraphics[scale=0.5]{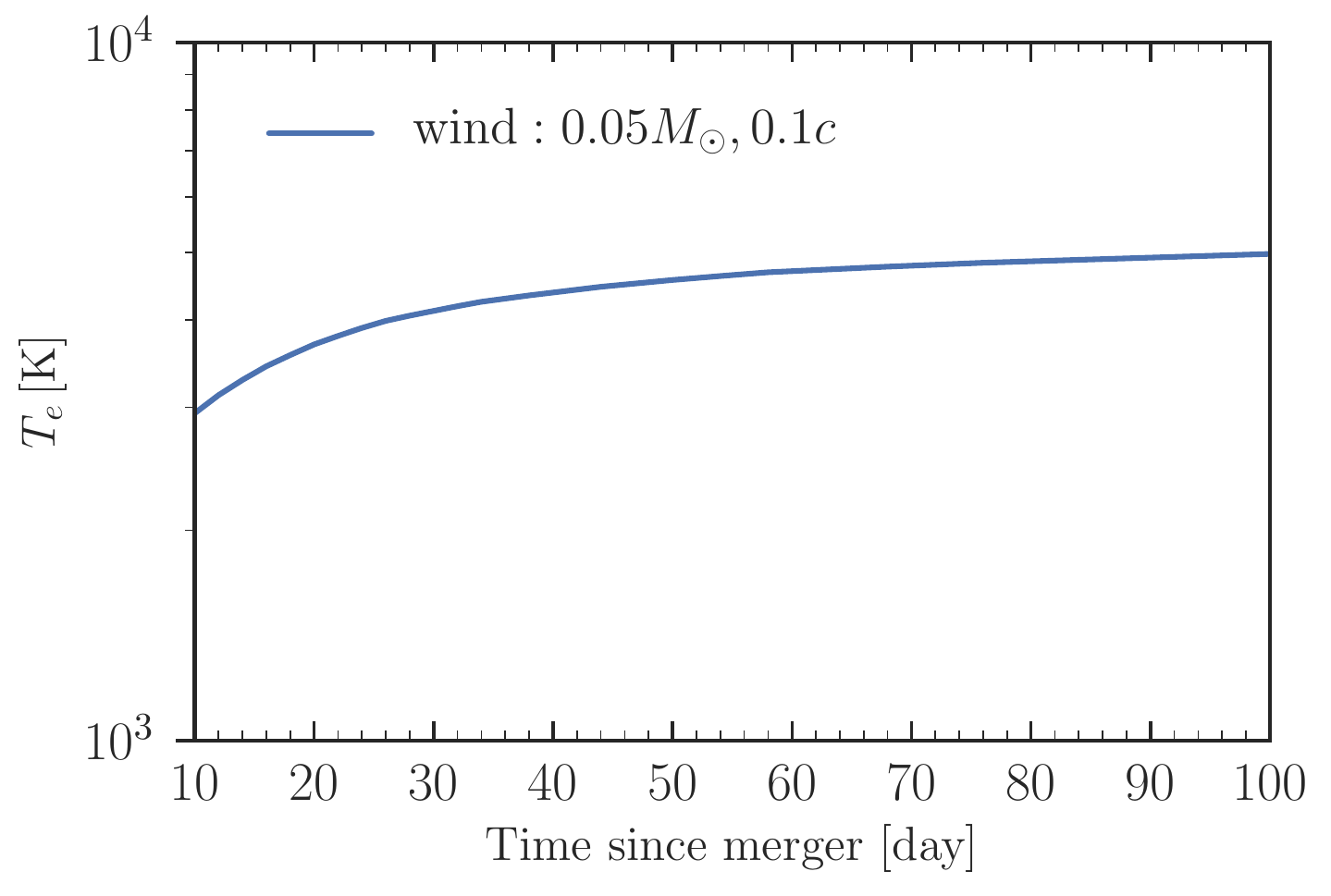}
\end{center}
\caption{
Evolution of fractional ion abundances, Nd I -- Nd IV ({\it left}) and kinetic temperature ({\it right}) for the fiducial model. The ionization degree and kinetic temperature increase until the thermalization time $t_{\rm th}$ and then become roughly constant with  time.
}
\label{fig:evolve}
\end{figure*}

\begin{table}
\caption{Model parameters.}
\label{tab:model}
\begin{center}
\begin{tabular}{lcc}
\hline \hline
model & $M_{\rm ej}\,[M_{\odot}]$ & $v_0\,[c]$ \\ \hline
 wind (fiducial) & $0.05$ &$0.1$ \\
 dynamical ejecta & $0.02$ & $0.2$ \\
 slow wind & $0.05$ & $0.05$ \\
\hline \hline\\
\end{tabular}
\end{center}
\end{table}

\section{Evolution of thermodynamic quantities and emergent spectrum}\label{sec:evolve}
By solving the  equations described in \S \ref{sec:eqs} with the atomic data of Nd ions shown in \S \ref{sec:atom}, we obtain the evolution of the thermodynamic quantities and emergent spectrum in the NSM nebular phase (see Appendix \ref{app:Ia} for an application of our method to SN Ia nebulae). Table \ref{tab:model} shows the three cases studied here and  we choose the wind model, $(M_{\rm ej},v_0)=(0.05M_{\odot},0.1c)$, as the fiducial model.

\begin{figure*}
\begin{center}
\includegraphics[scale=0.5]{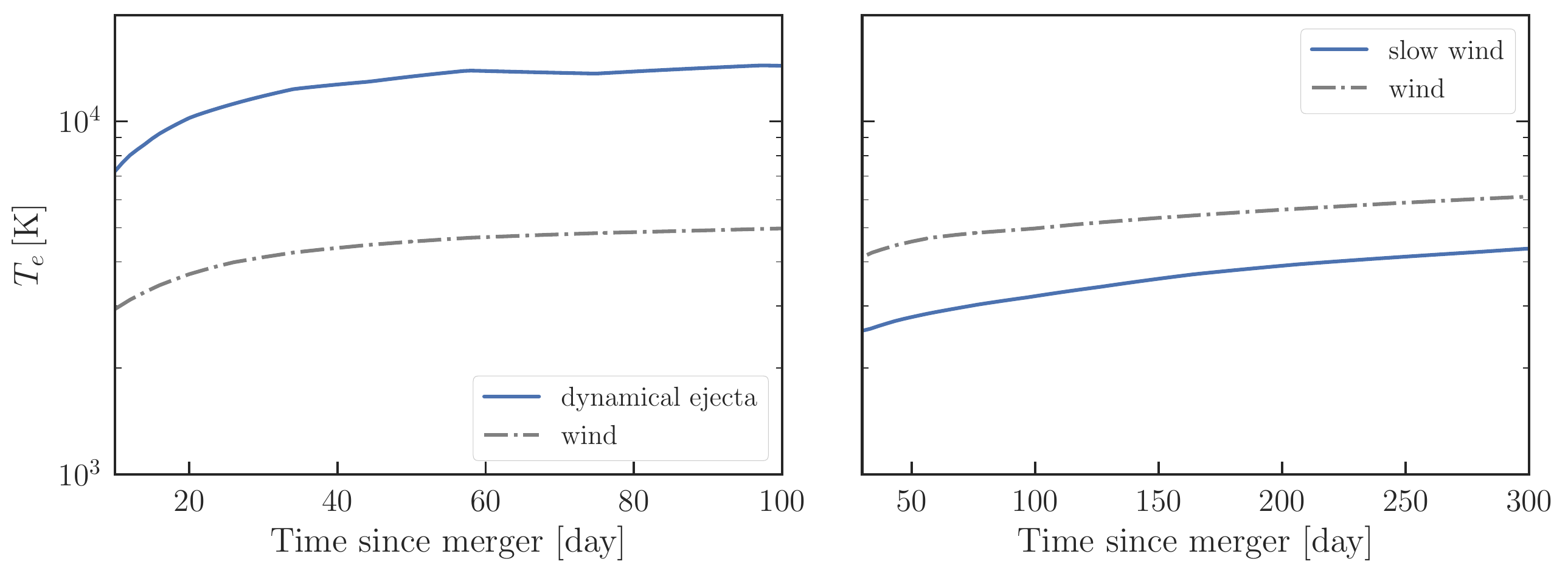}
\end{center}
\caption{Kinetic temperature evolution for the dynamical ejecta model (${\it left}$: $M_{\rm ej}=0.02M_{\odot}$ and $v_0=0.2c$) and slow wind model (${\it right}$: $M_{\rm ej}=0.05M_{\odot}$ and $v_0=0.05c$). The fiducial model (wind) is also shown as a dash-dotted curve for comparison.
The time scales on which the ejecta enters the nebular phase for the dynamical ejecta and slow models are $\approx 10\,{\rm day}$ and $70\,{\rm day}$, respectively.}
\label{fig:Tdy}
\end{figure*}

Figure \ref{fig:evolve} shows the evolution of  the fractional ion abundances and the kinetic temperature in the fiducial case.
The temperature  slowly increases with time from $\sim 3000\,{\rm K}$ to $5000\,{\rm K}$. {We find that the ejecta is predominantly composed of Nd II and Nd III.} 
As we discussed in \S \ref{sec:heat},
the evolution of these quantities becomes flat around the thermalization time, $t_{\rm th}\approx 50$\,day, where the normalized 
heating function changes its slope from $\propto t^{1.7}$ to $t^{0.2}$.
The fractional ion abundances also very slowly change with time after the thermalization break.  
\begin{figure*}
\begin{center}
\includegraphics[scale=0.5]{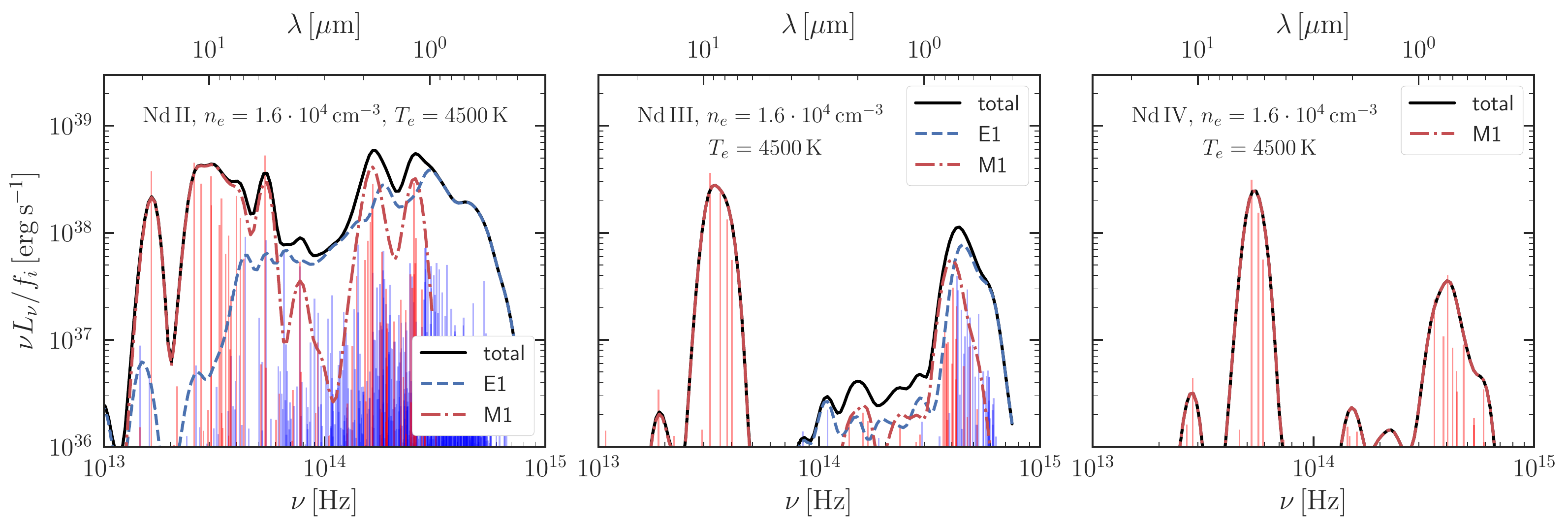}
\end{center}
\caption{
Normalized  spectra for  Nd II, Nd III, and Nd IV. Here we use a kinetic temperature of $T_e=4500\,{\rm K}$, an electron density of $n_e=1.6\cdot 10^4\,{\rm cm^{-3}}$, and electron fraction of $\chi=1$. These values roughly correspond to those around $40\,{\rm day}$ after merger in the fiducial model. Solid, dashed, and dash-dotted curves depict the total spectrum, the contribution of E1 transitions, and the contribution of M1 transitions, respectively. Also shown as vertical lines are individual E1 (blue) and M1 (red) lines.
The Doppler broadening of each line at a frequency $\nu_i$ is incorporated by using a Gaussian distribution with a standard deviation of $\approx (v_0/c)\nu_i=0.1\nu_i$.  
}
\label{fig:spec1}
\end{figure*}

\begin{figure*}
\begin{center}
\includegraphics[scale=0.5]{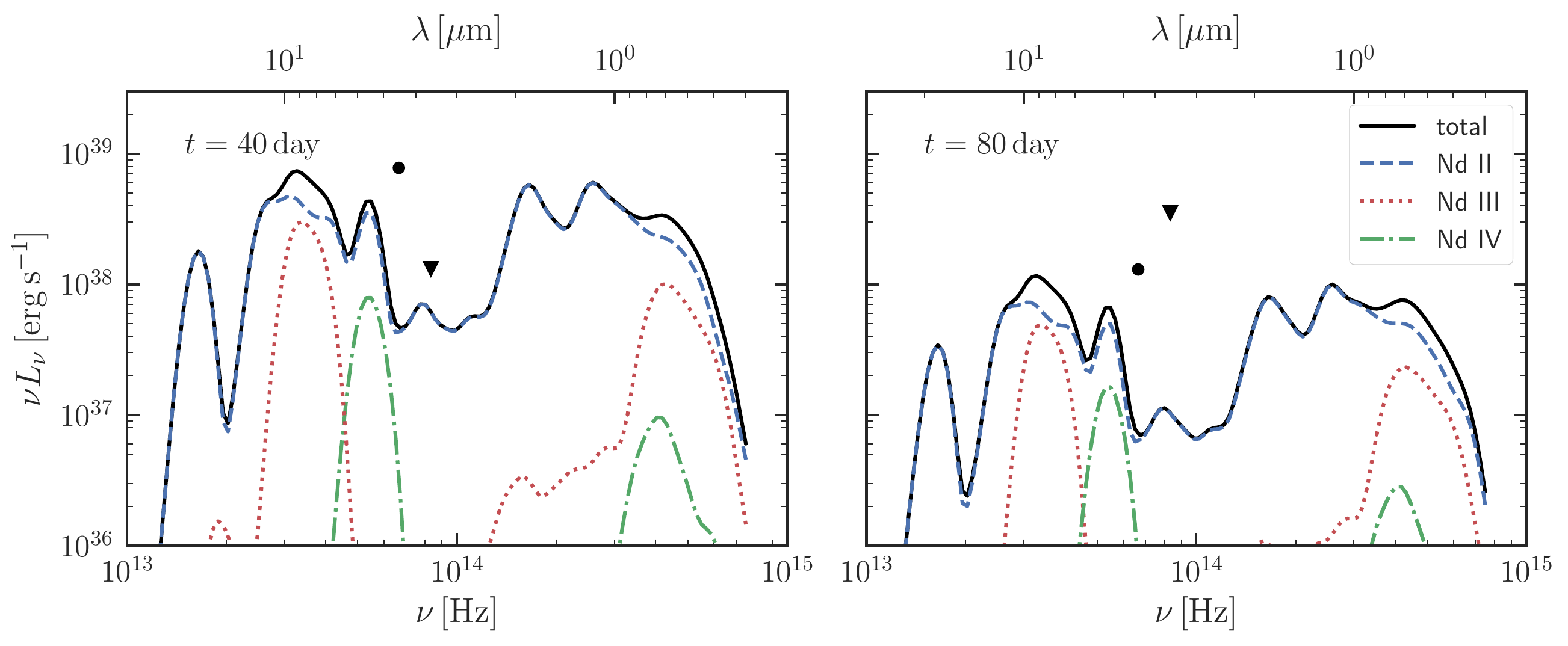}
\end{center}
\caption{Spectra for the fiducial model at 40 day (${\it left}$) and 80 day (${\it right}$) after merger. The contributions of Nd II, Nd III, Nd IV are also shown. Filled circle and triangle are the detection at $4.5\,{\rm \mu m}$ and $3\sigma$ upper limit at $3.6\,{\rm \mu m}$ obtained by {\it Spitzer} telescope at $43\,{\rm day}$ (${\it left}$) and $74\,{\rm day}$ (${\it right}$) after GW170817 \citep{Kasliwal2019MNRAS}.  
}
\label{fig:spec_ev}
\end{figure*}

Figure \ref{fig:Tdy} shows the temperature evolution for the dynamical ejecta and slow wind models.  
The characteristic temperatures for the dynamical ejecta and slow wind models are $\approx 10^4\,{\rm K}$ and $3\cdot 10^3\,{\rm K}$, respectively. The ionization degrees of dynamical ejecta and slow wind models are higher and lower than the fiducial model, respectively.

The  individual spectra of Nd II -- IV at $n_i = n_e=1.6\cdot 10^4 \,{\rm cm^{-3}}$ and $T=4500\,{\rm K}$ are shown in figure \ref{fig:spec1}.
In the nIR and optical region, these spectra can be understood qualitatively according to the characteristic spectra of lanthanides discussed in \S \ref{sec:atom}. Namely, these ions have two distinct  peaks, one around $5$--$10\,{\rm \mu m}$ produced by fine structure transitions and another around optical-nIR region.  Nd II  has among the richest spectral structure and its luminosity per atom is the brightest. The dense emission line distribution and the Doppler broadening result in a continuum-like spectrum with some  structures.
 We find that the following transitions predominately produce the Nd II spectrum: 4f$^3$5d$^2\rightarrow$4f$^4$5d,\, 4f$^3$5d6s$\rightarrow$4f$^4$6s,\, 4f$^3$5d6s$\rightarrow$4f$^4$5d,\, 4f$^4$6p$\rightarrow$4f$^4$5d,\,4f$^4$6p$\rightarrow$4f$^4$6s, and 4f$^4$5d$\rightarrow$4f$^4$6s. 
The Nd III and Nd IV spectra are produced by the transitions: 4f$^3$5d$\rightarrow$4f$^4$ and 
4f$^4\rightarrow$4f$^4$ for Nd III and 4f$^3\rightarrow$4f$^3$ for Nd IV. 
Note that individual M1 lines are more pronounced at $\lambda \lesssim 1\,{\rm \mu m}$ because
the line population in this wavelength region is less dense. 

\begin{figure*}
\begin{center}
\includegraphics[scale=0.5]{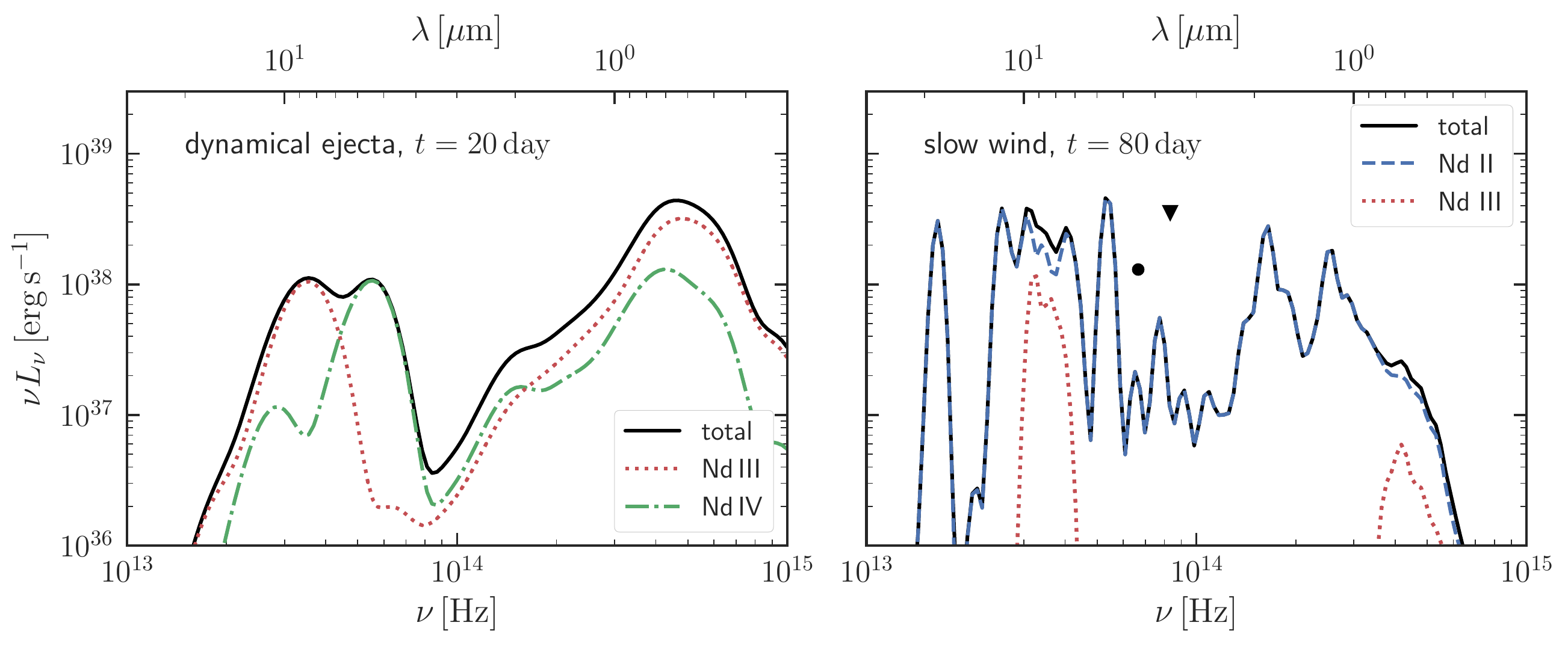}
\end{center}
\caption{
Same as figure \ref{fig:spec_ev} but for  the dynamical ejecta model at 20 day (${\it left}$) and the slow wind model at 80 day (${\it right}$).
}
\label{fig:spec_dy}
\end{figure*}

There are more E1 transition lines
at $\lambda \lesssim 1\,{\rm \mu m}$ for Nd II and Nd III (see figure \ref{fig:A}).
This implies that these E1 lines may absorb  other emission lines and reduce the emission at  $\lambda \lesssim 1\,{\rm \mu m}$.  
In fact, Nd II and Nd III respectively have $\sim 50$ and $\sim 15$  resonance lines in the range of $0.4\lesssim \lambda \lesssim 1\,{\rm \mu m}$  and $0.4\lesssim \lambda \lesssim 0.65\,{\rm \mu m}$. 
 This radiation transfer effect is not accounted for in our modeling, and thus, our modeling likely overpredicts the optical emission.

Figure \ref{fig:spec_ev} shows the total spectra at 40 and 80 day for the fiducial model with the fractional ion abundances shown in figure \ref{fig:evolve}. The Nd II lines dominate  the total spectrum particularly in the nIR band. The spectral shape does not change significantly from $40$ to $80\,{\rm day}$ while the amplitude decreases by a factor of $\sim 10$. This freeze-out of the nebular spectrum is a characteristic feature of the NSM nebular emission.

The spectra of the dynamical ejecta and slow wind models are shown in figure \ref{fig:spec_dy}. For dynamical ejecta,
each line is significantly broaden because of the fast expansion velocity, $0.2c$. As a result, the structures are completely smeared out. Nevertheless, there are two distinct peaks around the optical and IR bands. 
On the contrary, for the slow wind model, more lines can be seen in the IR region ($1\lesssim \lambda\lesssim 20\,{\rm \mu m}$) and the optical emission is very weak. The spectral shape does not evolve significantly during the nebular phase in the both models.

\begin{figure}
\begin{center}
\includegraphics[scale=0.5]{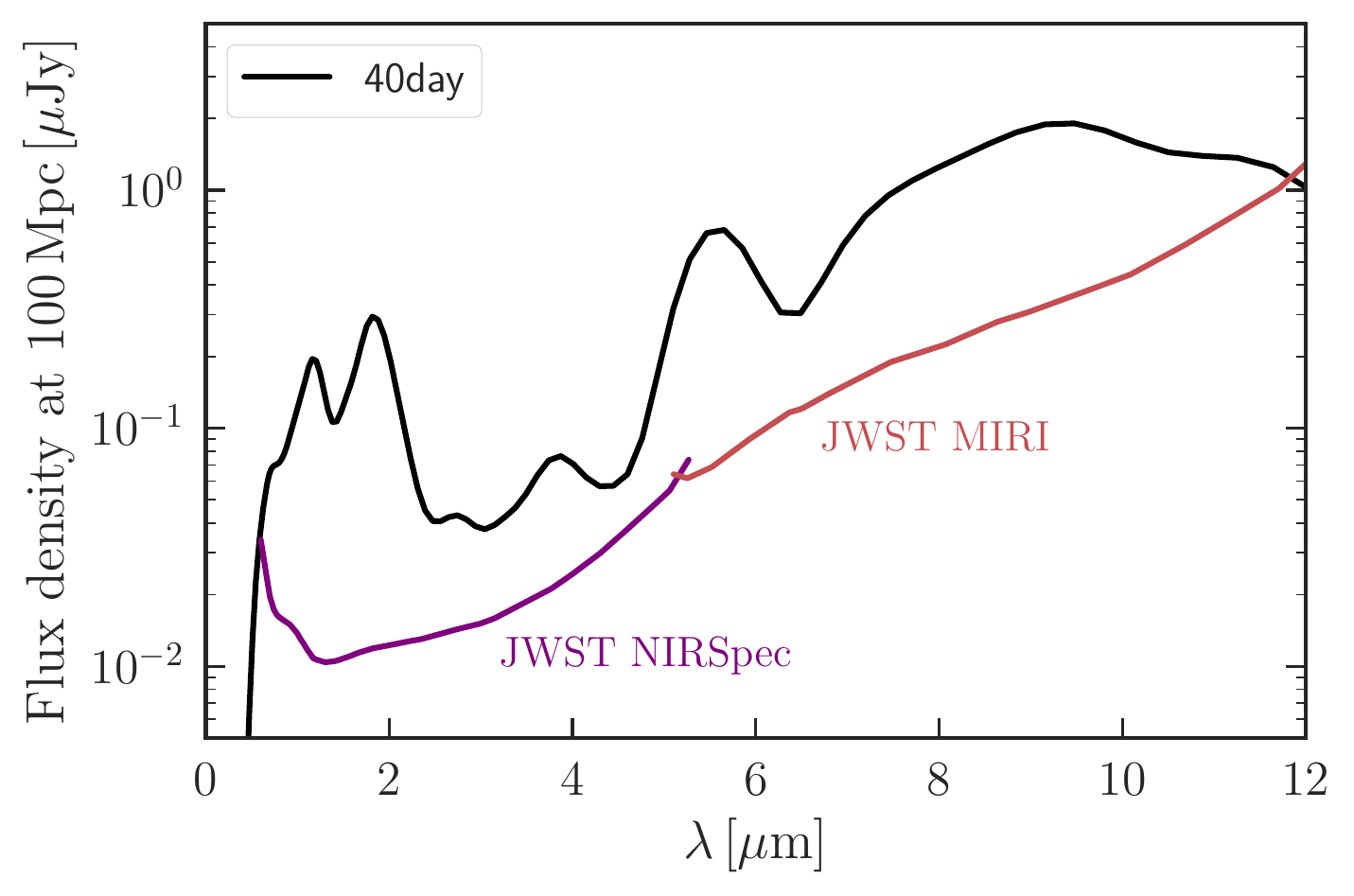}
\end{center}
\caption{ Spectrum at 40 day for the fiducial model at a distance of $100\,{\rm Mpc}$. Also depicted are the $1\sigma$ sensitivity curves of NIRSpec FS and MIRI LRS with $10^4\,{\rm s}$ integration.
}
\label{fig:spec_JWST}
\end{figure}

{We show the detectability of the structure of the nebular spectrum by the {\it James Webb Space Telescope} ({\it JWST}) for a future kilonova event in figure \ref{fig:spec_JWST}. The {\it JWST} is promising to resolve the spectral structure of the nebular emission around $40\,{\rm day}$ for events out to $\sim 100\,{\rm Mpc}$.  }


\section{Conclusion and discussion}\label{sec:conc}
The emission-line nebular phase of the  NSM ejecta is studied by using  a one-zone nebula model under non-LTE, in which the ejecta is considered to be  composed of one of lanthanide elements, Nd.
The atomic data necessary for the  modeling are calculated by using the atomic structure codes, \texttt{GRASP2K} and \texttt{HULLAC}. 
We find that the kinetic temperature and ionization fraction are nearly constant with time after the thermalization break of the beta-decay heating rate. Consequently, the spectral shape of the emergent emission is also expected to be frozen after the break. 
For the ejecta parameters of $M_{\rm ej}=0.05M_{\odot}$ 
and $v_0=0.1c$, we show that Nd II and Nd III are the most abundant ions and 
the kinetic temperature approaches $\approx  5000\,{\rm K}$. 

{The high ionization efficiency of the $\beta$-decay heating rate results in a deviation in the ionization state from LTE. In particular, we find that the neutral fraction is significantly suppressed in the nebular phase. Although we do not account for the velocity distribution in this work, we speculate that this deviation can occur even at the earlier times, e.g., $\lesssim 1$ week,  in the outer  ejecta, where the expansion velocity is faster. Depending on the mass and velocity, this effect leads to either the enhancement or suppression of singly ionized lanthanides, which has a crucial impact on the ejecta opacity \citep{tanaka2020,Barnes2020} and affects the color evolution of kilonovae \citep{Kawaguchi2020}.   }

The emergent emission line spectrum of the pure Nd nebula  consists of a broad structure from $\sim 0.5{\rm \mu m}$ to $20{\rm \mu m}$ with two distinct peaks around $\sim 1\,{\rm \mu m}$ and $\sim 10\,{\rm \mu m}$.   Fine-structure transitions produce the mid-IR peak. This spectral structure may be an unique feature of lanthanide-rich nebulae. It is worth emphasizing that individual M1 lines are more pronounced at $\lambda \lesssim 1\,{\rm \mu m}$ because
the line population in this wavelength region is less dense. 
Importantly, the {\it JWST} will be able to  resolve such structure in the nIR and midIR regions  for events  at $\sim 100\,{\rm Mpc}$. 
Note, however, that this structure may be suppressed once more elements are included.  Another caveat of our modelling is that we have neglected the absorption due to line overlapping, which may lead to an overestimate of the optical-nIR emission ($\lambda \lesssim 1\,{\rm \mu m}$), where 
Nd II and Nd III have a number of permitted lines.

We use a crude approximation for the collisional strength
of forbidden lines, i.e., $\Omega_{F}=1$, in the case of the \texttt{GRASP2K} calculation. While this approximation is statistically consistent with the collisional strengths derived with \texttt{HULLAC} and can reasonably reproduce the cooling rates, the predicted line intensity ratios are by no means accurate. Thus, we need more accurate collisional strengths for the future studies.

\section*{Acknowledgments}
We thank B. T. Draine, M. M. Kasliwal, K. Kawaguchi, and E. Nakar  for useful discussion and M. Busquet for the generous support on the \texttt{HULLAC} code. K. H. was supported by Japan Society for the Promotion of Science (JSPS)  Early-Career Scientists Grant Number 20K14513.

\section*{DATA AVAILABILITY}
The data underlying this article will be shared on reasonable request to the corresponding author.

\bibliographystyle{mnras}
\bibliography{ref,Rprocess}

\appendix

\section{Work per ion pair}\label{sec:w}
The ionization efficiency of fast electrons for a  stopping plasma is in principle obtained by solving the Boltzmann equation under some approximations \citep{Spencer1954PhRv,Kozma1992ApJ}. Here we take a simple approach employed by \cite{Axelrod1980}, which describes the radioactive ionization rate in terms of work per ion pair.   {The work per ion pair of $X^{i+}$ is defined by}
\begin{eqnarray}
w_i = \frac{f_i E_{\rm dis}}{N_i},\label{eq:w}
\end{eqnarray}
{where $f_i$ is the number fraction $X^{i+}$, $E_{\rm dis}$ is the total dissipated energy of injected fast elections  and $N_i$ is the total number of ion pairs (ion-electron pairs) of $X^{(i+1)+}$ produced through by the fast electrons. The value of $w_i$ simply represents the amount of energy that is dissipated in each ion-electron pair production.}

{Let us consider first the work per  ion pair for a primary electron  with an initial kinetic energy of $E_p$ injected in a stopping plasma.
The number of ion pairs of $X^{(i+1)+}$ through the thermalization of the primary is given by}
\begin{eqnarray}
N_{i,p} = \int_0^{s_{\rm th}} n_i \sigma_i(E(s)) ds,\label{eq:Np1}
\end{eqnarray}
where $n_i$ is the number density of $X^{i+}$,
$s$ is the travel distance of the electron, and $\sigma_i$ is the ionization cross section. The energy loss per  distance interval is 
\begin{eqnarray}
\frac{dE}{ds} \approx -n\left(L(E)+L_{\rm th}(E,\chi)\right),
\end{eqnarray}
where $L$ and $L_{\rm th}$ are
 the stopping cross sections due to collisional ionization and excitation,  and due to the Coulomb collision with thermal electrons, respectively. Equation (\ref{eq:Np1}) is rewritten as
\begin{eqnarray}
N_{i,p} = f_i\int_{kT_e}^{E_p}  \frac{\sigma_i(E)}{L(E)+L_{\rm th}(E,\chi)} dE.\label{eq:Np}
\end{eqnarray}
The work per ion pair of the primary electron is then 
\begin{eqnarray}
w_{i}^{p} = \frac{E_p}{\int_{0}^{E_p} dE \sigma_{i}/(L(E)+L_{\rm th}(E,\chi))}.\label{eq:wp}
\end{eqnarray}

To evaluate equation (\ref{eq:wp}), we use the total ionization cross section of $X^{i+}$ by electron-ion collision  given by \cite{Axelrod1980}
\begin{eqnarray}
\sigma_i \approx \frac{2\pi e^4}{mv^2} \sum_{j=1}^{N}\frac{q_j}{P_j} \left[\ln\left(\frac{ m_ev^2}{2P_j}\right) -\ln(1-\beta^2)-
\beta^2 \right],\label{eq:cross}
\end{eqnarray}
where $v$ is the electron's velocity,
$\beta=v/c$, $q_j$ and $P_j$  are the number of electrons and the ionization potential of a subshell $j$. 
We note that this formula (\ref{eq:cross})  agrees with the experimental data \citep{yagi2001}. The stopping power for electrons is given by the Bethe formula:
\begin{eqnarray}
L(E) & \approx & \frac{4\pi Ze^4}{m_e v^2} \left[\ln\left(\frac{\sqrt{m_ev^2T}}{2^{1/2}\langle I\rangle(1-\beta^2)^{1/2}} \right)\right. \\
  - (\sqrt{1-\beta^2} &-& \left.\frac{1-\beta^2}{2})\ln2 +\frac{1-\beta^2}{2} +\frac{1}{16}(1-\sqrt{1-\beta^2})^2\right],\nonumber \label{eq:ele}
\end{eqnarray}
where $Z$ is the charge of the target ion, $T$ is the kinetic energy of the electron, and
$\langle I \rangle$ is the mean ionization energy of the stopping material. The value of $\langle I \rangle$ is taken from the ESTAR database\footnote{https://physics.nist.gov/PhysRefData/Star/Text/ESTAR.html}.
The stopping power  of thermal plasma for with thermal velocity $v_{\rm th}\ll v$ is given by \citep{bohr}
\begin{eqnarray}
L_{\rm th}(E,\chi) \approx \frac{4\pi e^2 \chi}{m_e v^2}\ln\left(\frac{1.123m_ev^3}{e^2\omega_p} \right),
\end{eqnarray}
where $\omega_p$ is the plasma frequency, $n_e$ is the electron number density, and $\chi$ is the electron fraction $n_e/n$. 

 
 For comparison between different ions, it is useful to define  work per  ion pair normalized by the first ionization potential, $I_{i,1}$.
For instance, in the case of $E_p = 250\,{\rm keV}$ and $\chi=2$, we find $w_{i}^p/I_{i,1} \sim 45$, $45$, $40$, and $35$ for Nd\,I, Nd\,II, Nd\,III, and Nd\, IV, respectively. In addition to  ionization by primary electrons, secondary electrons may cause further ionization. 
This means that the total number of ion pairs $N_{i}$ in equation (\ref{eq:w}) is larger than $N_{i,p}$.
Secondaries  are ejected with recoil energy typically around the binding energy of the target electron. 
For a weakly ionized Nd plasma, the stopping power of thermal electrons dominates over the ionization energy loss at electron energies $\lesssim \langle I \rangle\approx 0.5$ keV, and therefore, the recoil energy of the secondaries originating from the inner shells (K, L, M) typically exceeds this threshold. Thus, a  fraction of the recoil energy of  secondaries from the inner shells is lost through ionization and  more ion pairs are created.  Accounting for the secondary ionization, the ionization efficiency is increased by $\sim 20$--$50\%$ corresponding to  $w_i/I_{i,1}\sim 30$. In this paper, we use $w_i/I_{i,1}= 30$.

\section{Photoionization}\label{app:photo}
The recombination processes emit photons that may be reprocessed by photoelectric absorption. This reprocess reduces the recombination rate.   The recombination of $X^{j+}$ may emit  
ionizing photons for $X^{(i<j)+}$. 
The number of ion pairs of $X^{(j+1)+}$ and $e^-$ produced by photoionization due to the photons emitted in a recombination process of $X^{(j+1)+}\rightarrow X^{j+}$  is estimated by
\begin{eqnarray}
P_{ij} = \int \left(1-e^{-\tau}\right)\frac{f_i\sigma_{i}^{\rm ph}}{\sum_k f_k\sigma_{k}^{\rm ph}}\left(\frac{dN_{\rm ph}}{d\nu} \right)_j d\nu,
\end{eqnarray}
where $(dN_{\rm ph}/d\nu)_j$ is the number of photons per frequency interval emitted in recombination of $X^{(j+1)+}$. Here
the optical depth for photons  with frequency $\nu$  is given by 
\begin{eqnarray}
\tau(\nu) & \approx & \sum_{k=0} f_k\sigma_{k}^{\rm ph}(\nu)n R,
\end{eqnarray}
where $f_k$ and $\sigma_{k}^{\rm ph}$ are the number fraction and the photoionization cross section of $X^{k+}$, and $R$ is the radius of the ejecta. Because $\sigma^{\rm ph}$ is $\mathcal{O}(10^{-18}\,{\rm cm^2})$, the optical depth is $\sim 100\,(t/40\,{\rm day})^{-2}$ and therefore the ejecta is optically thick for recombination photons in the nebular phase. 

As shown in figure \ref{fig:recombination}, Nd ions recombine predominantly through dielectronic recombination, where recombination photons are produced through the radiative cascade of auto-ionization states to the ground state.   Therefore, it is not straightforward to determine the recombination photon spectrum $(dN_{\rm ph}/d\nu)_j$. For auto-ionization states that have a large radiative transition rate, each auto-ionization  state contributes substantially to the rate coefficient even though the number of such states is relatively small. In this channel, auto-ionization states are typically stabilized through the emission of a photon with energy close to the first ionization potential of the recombined ion and therefore this cascade produces one ionizing photon and several low energy photons. At the same time, there are many auto-ionization states that are stabilized through the emission of photons with energy sufficiently lower than the first ionization potential but high enough to ionize ions in lower ionized states.  As a result, the recombination photons are likely to have a somewhat flat spectrum per logarithmic frequency interval.   Thus, we assume that the number of recombination photons is constant at each energy scale below the sum of the first ionization potential and the thermal energy of free electrons, i.e., $(dN_{\rm ph}/d\nu)_j\propto 1/\nu$ for $h\nu<I_{j,1}+kT_e$ and  its normalization is set such that the total energy of recombination photons is $I_{j,1}+kT_e$.



\section{Self-absorption of strong lines}\label{app:esc}
The absorption due to strong lines may have significant impacts on the cooling functions and emergent spectra. In general, absorption occurs non-locally so that one must solve radiation transfer, which is beyond the framework of our one-zone modeling. Here we use the escape probability approximation,  which allows to include the effects of self-absorption of  lines in one-zone modelings (e.g, Chapter 19 of \citealt{Draine}). 

In  homologously expanding ejecta,
the escape probability is approximated by 
\begin{eqnarray}
\langle \beta_{ij} \rangle = \frac{1-e^{-\tau_{s,ij}}}{\tau_{s,ij}},
\end{eqnarray}
where $\tau_{s,ij}$ is the Sobolev optical depth:
\begin{eqnarray}
\tau_{s,ij} &=& \frac{g_iA_{ij}}{8\pi} \lambda_{ij}^3 \left( \frac{n_j}{g_j}-\frac{n_i}{g_i}\right)t~~~~~(i>j).
\end{eqnarray}
For resonance lines, the optical depth is estimated by
\begin{eqnarray}
\tau_{s,{\rm res}}  & \sim & 10^2~\left(\frac{g_uA_{u0}}{10^{6}\,{\rm s^{-1}}} \right)\left(\frac{\lambda_{u0}}{0.5\,{\rm \mu m}} \right)^{3}
\left(\frac{M_{\rm ej}}{0.05\,M_{\odot}} \right)\nonumber\\ 
& & \times\left(\frac{v_0}{0.1c} \right)^{-3}
\left(\frac{t}{30\,{\rm day}} \right)^{-2}.
\end{eqnarray}
This suggests that the resonance lines are trapped in the ejecta on time scales focused in this paper, $\lesssim 100$ day. 
 

\begin{figure*}
\begin{center}
\includegraphics[scale=0.5]{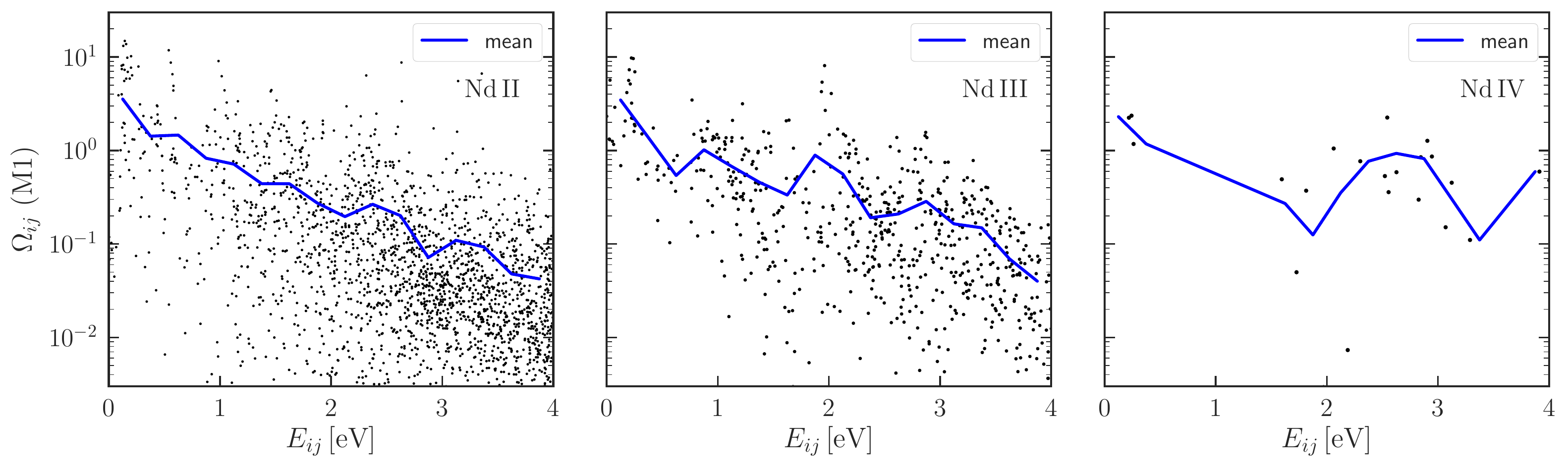}
\end{center}
\caption{
Velocity averaged collision strengths for M1 transitions at $T_e=5000\,{\rm K}$ as a function of the transition energy computed by using \texttt{HULLAC}. The averaged value is shown by a blue line. 
}
\label{fig:coll}
\end{figure*}

\begin{figure*}
\begin{center}
\includegraphics[scale=0.5]{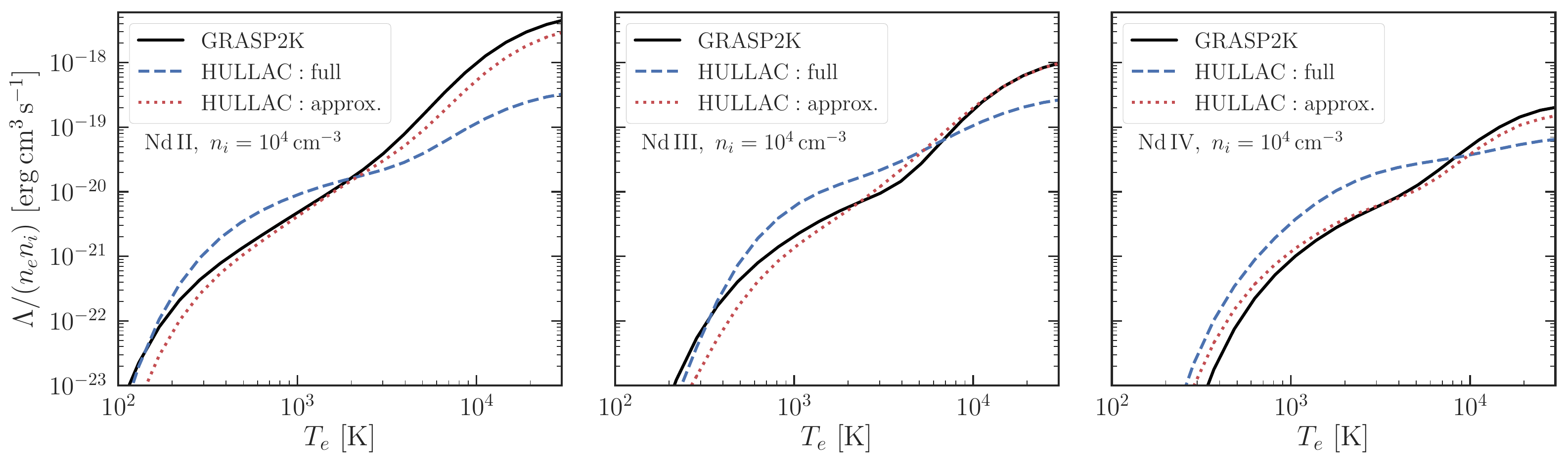}
\end{center}
\caption{
Cooling functions for different atomic data. {\it Solid curve} shows the cooling function for the \texttt{GRASP2K} atomic data with the formula (\ref{eq:vR}) and $\Omega_F=1$. Also shown are the cooling functions derived the \texttt{HULLAC} atomic data, where collisional strengths are computed ({\it dashed curve}).  {\it Dotted curve} shows the cooling function with the \texttt{HULLAC} energy levels and $A_{ij}$ but collisional strengths are approximated by  the formula (\ref{eq:vR}) and $\Omega_F=1$.
}
\label{fig:chullac}
\end{figure*}

\begin{figure*}
\begin{center}
\includegraphics[scale=0.5]{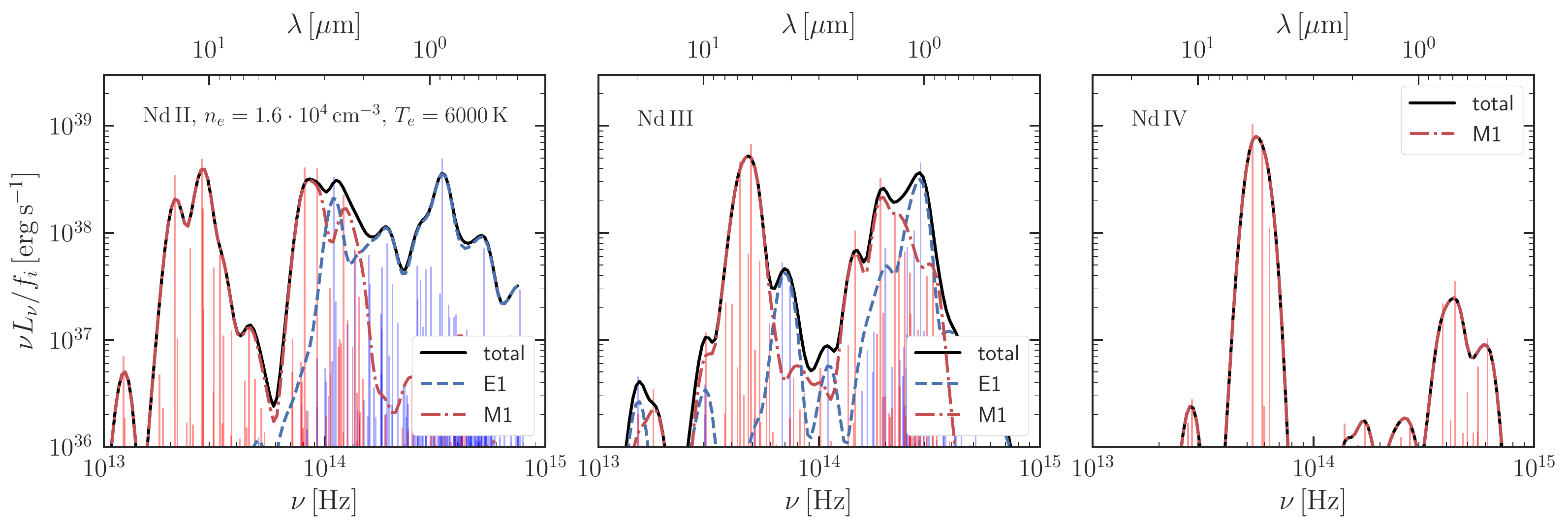}
\end{center}
\caption{
Same as figure \ref{fig:spec1} but the energy levels, radiative transition rates, and collisional strengths are computed with \texttt{HULLAC}.
Here we set $T_e=6000\,{\rm K}$ instead of $4500\,{\rm K}$ because the temperature required from the heating rate around $40\,{\rm day}$ is slightly higher than that of \texttt{GRASP2K} due to the low cooling function of Nd II of \texttt{HULLAC}.}
\label{fig:spechullac}
\end{figure*}


\section{Collisional excitation and deexcitation}\label{app:coll}
With the usual convention,
the velocity averaged rate coefficient for collisional deexitation  from an upper level $i$ to a lower level $j$ is given by
\begin{eqnarray}
k_{ij} = \frac{8.63\cdot 10^{-6}\Omega_{ij}(T_e)}{g_iT_e^{1/2}}\,{\rm cm^3s^{-1}},
\end{eqnarray}
where $\Omega_{ij}$ is the velocity averaged collision strength connecting levels $i$ and $j$. 
The collisional excitation rate coefficient is given by
\begin{eqnarray}
k_{ji} = \frac{g_i}{g_j}k_{ij}e^{-E_{ij}/kT_e},
\end{eqnarray}
where $g_{i}$ is the level degeneracy, $E_{ij}$ is the energy-level difference. 

The collisional strengths are currently not available for the \texttt{GRASP2K} atomic data. Therefore, we use the following approximations for the collisional strengths $\Omega_{ij}$ for  the \texttt{GRASP2K} atomic data. 
For E1 transitions, we  calculate $\Omega_{ij}$ by using the approximate formula  \citep{vanRegemorter1962ApJ}:
\begin{eqnarray}
\Omega_{ij} \approx 2.388 P\left( \frac{E_{ij}}{kT_e}\right) \left(\frac{\lambda}{1\,{\rm \mu m}}\right)^{3}
\left(\frac{g_i A_{ij}}{10^6\,{s^{-1}}}\right),\label{eq:vR}
\end{eqnarray}
where $P(x)$ is the Gaunt factor integrated over the electron velocity distribution and $A_{ij}$ is computed with \texttt{GRASP2K}. Here we use $P(x)\approx 0.2$, which is a good approximation for $x\lesssim 2$ \citep{vanRegemorter1962ApJ}.
For forbidden transitions, we  assume $\Omega_{ij}=\Omega_F$, where $\Omega_F$ is a constant value. 
Figure \ref{fig:coll} shows the collisional strengths for M1 transitions at $T_e=5000\,{\rm K}$ computed by using \texttt{HULLAC}. We find that the averaged values around $E_{ij}\lesssim 1\,{\rm eV}$, which are the most relevant to the spectrum formation in the nebular phase,  is roughly unity.
 Therefore, we approximate $\Omega_F\approx 1$ in this work. 
Figure \ref{fig:chullac} compares the cooling function of \texttt{GRASP2K} with that of  \texttt{HULLAC}. The cooling functions due to M1 transitions derived with the two codes are in a good agreement. This fact justifies our choice of $\Omega_F\approx 1$.  However, the E1 transition cooling of \texttt{GRASP2K} is higher than that of \texttt{HULLAC}, suggesting that the van Regemorter formula slightly overestimates the collisional strengths. 
 Figure \ref{fig:spechullac} shows the spectrum of each ion at $n=1.6\cdot 10^{4}\,{\rm cm^{-3}}$ and  $T_e=6000\,{\rm K}$ with the atomic data computed with \texttt{HULLAC}. 
We note that the spectral structures computed with the two codes are qualitatively similar but the Nd II spectrum of \texttt{HULLAC} has  significant emission around $3\,{\rm \mu m}$.

\section{Dielectronic recombination}\label{app:rec}
Dielectronic recombination occurs via the following process:
\begin{eqnarray}
X_{p}^{(i+1)+} + e^{-} & \rightarrow & X_{a}^{i+}
 \rightarrow  X_{b}^{i+}+h\nu,\label{eq:cap}
\end{eqnarray}
where 
$a$ and $b$ denote an autoionizing state of $X^{i+}$ and a bound state of $X^{i+}$, respectively. The bound state, $X_{b}^{i+}$,  is stabilized by radiative decays.  At the nebular temperature, radiative decays of both the core and captured electrons contribute to the stabilization of $X^{i+}_a$ \citep{Beigman1980JPhB,Storey1981MNRAS}.  


The dielectronic recombination rate coefficient of the capture process (\ref{eq:cap}) is calculated by
\begin{eqnarray}
\alpha_{\rm di}(p,a;T_e) = \left(\frac{N_S(X_{a}^{i+})}{N_eN_S(X_{p}^{(i+1)+})} \right)
 \frac{\sum_j A_{aj}\sum_c\Gamma_{ac}}{\sum_c \Gamma_{ac}+\sum_{k}A_{ak}},
\end{eqnarray}
where 
\begin{eqnarray}
\frac{N_S(X_{a}^{i+})}{N_eN_S(X_{p}^{(i+1)+})} = \frac{g_a}{2g_{p}}
\left(\frac{h^2}{2\pi m_e kT_e} \right)^{3/2}e^{-E_a/kT_e},
\end{eqnarray}
where $E_a$ is the energy of a state $a$ relative to $X_p^{(i+1)+}$,
and $g_p$ is the statistical weight of the state $p$,
$\Gamma_{ac}$ is the autoionization rate. Here the sum for $j$ is over the levels that are stable against autoionization and the sum for $k$ is over all the lower levels.  
We assume that the recombining ion $X_{p}^{(i+1)+}$ is in the ground state. Then the total rate coefficient is 
\begin{eqnarray}
\alpha_{\rm di}^{\rm tot}(T_e) = \sum_a\alpha_{\rm di}(p,a;T_e).
\end{eqnarray}

This capture is a resonant process such that $E(e^-)=E(X^{i+}_a)-E(X^{(i+1)+}_p)$ must be satisfied and the autoinizing states that are accessible via collision with thermal electrons contribute to the capture rate. 
This indicates that ions with  denser autoionizing states such as open f-shell ions have larger recombination rate coefficients.
In fact,  the  measured values of the dielectronic recombination rate coefficient of Au$^{25+}$, W$^{20+}$, and W$^{18+}$, nearly half open f-shell ions, are larger than the radiative recombination rate coefficient  by two to three orders of magnitude at nebular temperatures \citep{Hoffknecht_1998,Schippers,Spruck2014PhRvA}.

For the nebular temperatures ($\lesssim 10^4\,{\rm K}$), the kinetic energy of thermal electrons is typically much smaller than the first ionization potential of ions, and therefore, autoionizing states only slightly above  the ionization threshold contribute to the recombination process. Thus,  levels in a small energy range from $I_1$ to $\sim I_1 + 1\,{\rm eV}$ must be resolved.
For this purpose, we use the level mode of \texttt{HULLAC} that resolves  the fine structure. 

\section{Radiative recombination}
Radiative (direct) recombination occurs via
\begin{eqnarray}
X_{p}^{(i+1)+} + e^{-}  \rightarrow X_{b}^{i+}+h\nu.\label{eq:cap_r}
\end{eqnarray}
A photon produced by the recombination of $X^{i+}$ directly to the ground state is most likely absorbed by $X^{i+}$. This rate coefficient of this process is denoted customary as $\alpha_A$ and that of the recombination to the other states is denoted $\alpha_B$. \cite{Axelrod1980} provides
\begin{eqnarray}
\alpha_A(T) = 10^{-13}i^2\left(\frac{T}{10^4\,{\rm K}}\right)^{-1/2}\,{\rm cm^3\,s^{-1}},
\end{eqnarray}
and 
\begin{eqnarray}
\alpha_B(T) = 3\cdot 10^{-13}i^2\left[\left(\frac{T}{10^4\,{\rm K}}\right)^{-3/2}-\frac{1}{3}\left(\frac{T}{10^4\,{\rm K}}\right)^{-1/2}\right]\,{\rm cm^3\,s^{-1}}.\label{eq:B}
\end{eqnarray}
This form is provided for iron but it is not significantly different for heavy elements.
We include the case B recombination (equation \ref{eq:B}) in our modeling.

\begin{figure}
\begin{center}
\includegraphics[scale=0.3]{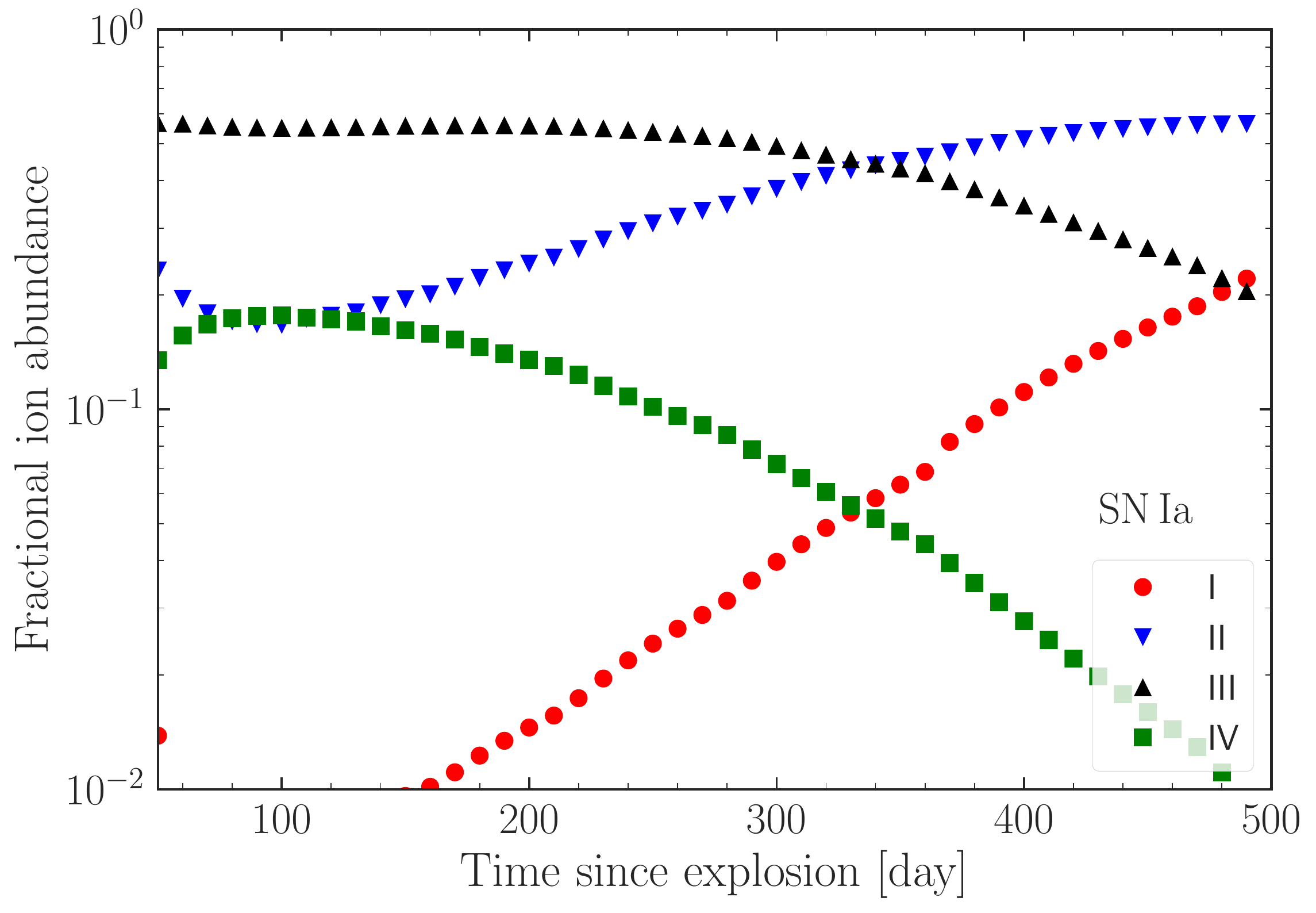}
\end{center}
\caption{
Evolution of fractional ion abundances of Fe I -- IV. Here we use $M_{\rm ^{56}Ni}=0.54M_{\odot}$
and $v_0=7000\,{\rm km\,s^{-1}}$. 
}
\label{fig:ia}
\end{figure}

\begin{figure}
\begin{center}
\includegraphics[scale=0.5]{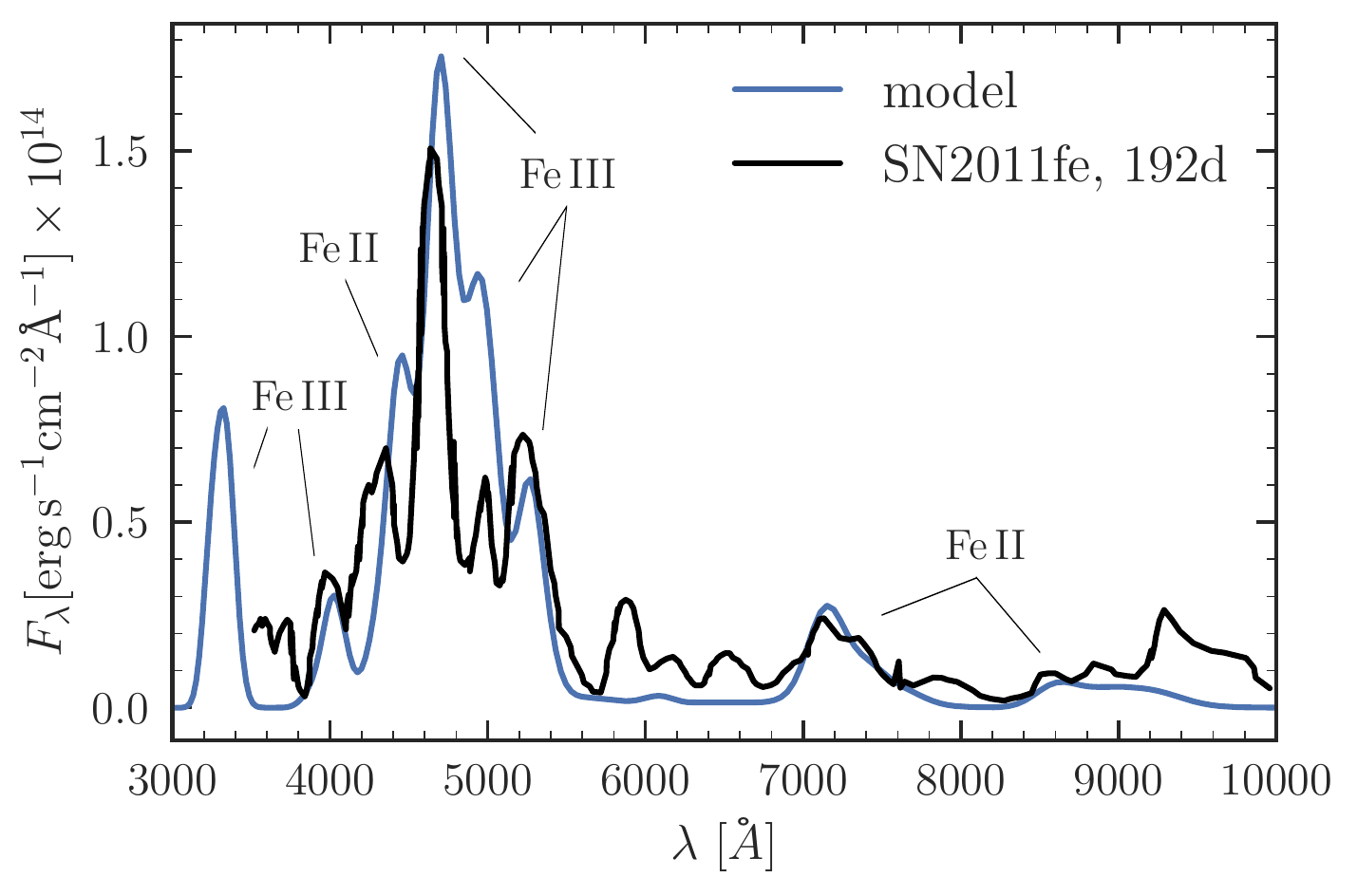}
\includegraphics[scale=0.5]{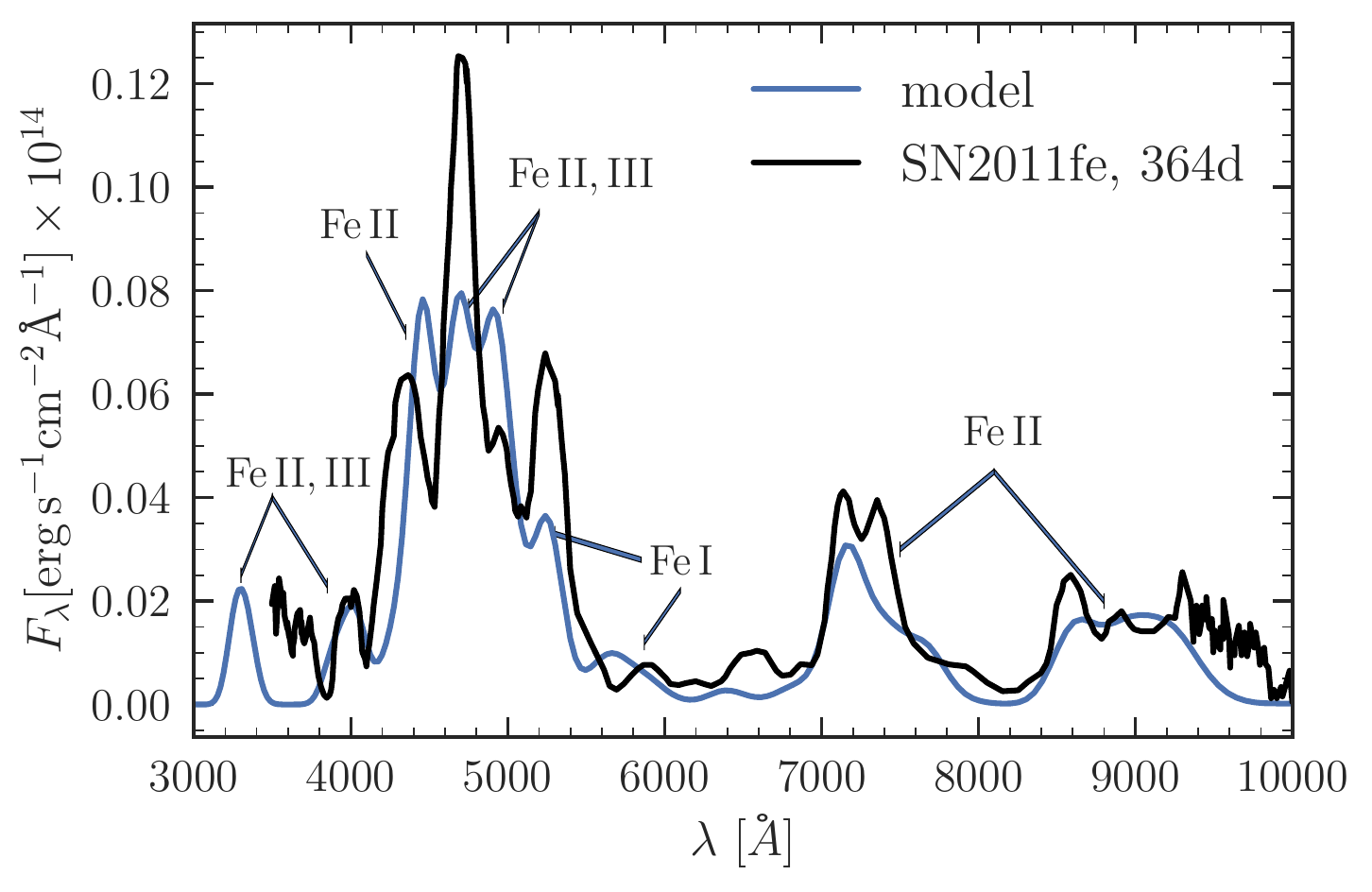}
\end{center}
\caption{
Spectra of a pure Fe nebula. Here we use $M_{\rm ^{56}Ni}=0.54M_{\odot}$
and $v_0=7000\,{\rm km\,s^{-1}}$. $\Omega_{F}=0.5$ with which the cooling functions computed with \texttt{HULLAC} and 
the simple approximated method.
Also shown is the observed nebular spectrum of SN2011fe \citep{Mazzali2015MNRAS}. Because our model includes only Fe ions some lines are absent in the synthetic spectra, e.g., Co\,III lines around $6000{\AA}$.
}
\label{fig:ia_s}
\end{figure}

\section{Nebular spectra of SNe Ia}\label{app:Ia}
Our nebula modeling is by no means accurate because we use a number of approximations and  assumptions. In order to show the ability of our simple modeling, here
we apply our method to the nebular emission of SNe Ia for comparison. Here we consider the decay chain of $^{56}{\rm Ni}\rightarrow ^{56}{\rm Co} \rightarrow ^{56}{\rm Fe}$ as the heat and ionization source and the heating rate computed by a code developed by \cite{hotokezaka2020ApJ}. As we did for NSM nebulae, we assume that the atomic properties are represented by a single atomic species, Fe. The  work per ion pair for Fe ions for primary electrons is $\omega_p/I_1\approx 30$ \citep{Axelrod1980}.
Accounting for secondary ionization, we approximate $\omega/I_1 \approx 25$.
Because the properties of the transition lines of Fe ions relevant to the SN Ia nebula modelings are experimentally known, we use the NIST line list instead of preparing them with the atomic codes. The collisional strengths are computed in the prescription shown in Appendix \ref{app:coll}. Here we use $\Omega_{F}=0.5$ for forbidden transitions, with which the cooling functions agree with those computed by using \texttt{HULLAC} in the relevant temperature range.
Note that our one-zone modeling is fully characterized by only two parameters: the total $^{56}$Ni mass, $M_{^{56}{\rm Ni}}$, and  the ejecta velocity, $v_0$.

We discussed in \S \ref{sec:rec} that dielectronic recombination dominates over radiative recombination for Nd ions.  Likewise, dielectronic recombination is more important for lower ionized Fe ions (see \citealt{Nahar_feI,Nahar_feII,Nahar_feIII} for the results of the R-matrix method). We obtain the rate coefficients of dielectronic recombination of Fe ions by using \texttt{HULLAC}. We find that our rate coefficients
are higher by a factor of $\sim 3$--$10$ than those of \cite{Nahar_feI,Nahar_feII,Nahar_feIII}.

Figure \ref{fig:ia} shows the evolution of ionization fractions in the case of $M_{^{56}{\rm Ni}}=0.54M_{\odot}$ and $v_0=7000\,{\rm km\,s^{-1}}$. 
Figure \ref{fig:ia_s} shows the spectrum of pure Fe emission at $\sim 200\,{\rm day}$ and $\sim 360\,{\rm day}$. Also depicted is the observed spectrum of a typical SN Ia, SN\,2011fe \citep{Mazzali2015MNRAS}. Our simple one-zone model reproduces the characteristic Fe-line structure  (see \citealt{Mazzali2015MNRAS} and \citealt{Boty2018ApJ} for more detailed modelings).  Note that our choice of $M_{\rm ^{56}Ni}=0.54M_{\odot}$ agrees with the mass estimate by using the pre-nebular light curve of SN\,2011fe \citep{Arnett2017ApJ}. At $\sim 360$ day, the value of Fe\,III/Fe\,II in our model seems slightly lower than the observed value and the model predictions in the literature. This is because    our dielectronic recombination rate coefficients, which are computed with \texttt{HULLAC},  are slightly larger than those used in the literature.

\bsp	
\label{lastpage}
\end{document}